\documentclass[12pt]{article}

\usepackage{cite,amsmath,amsfonts,amsthm,fullpage}

\newcommand{\sgn}{\mathop\mathrm{sgn}\nolimits}

\def\bt{{\mathbf t}}
\usepackage{amssymb}
\usepackage{amsbsy}
\usepackage{epsfig}
\usepackage{verbatim}
\renewcommand{\theequation}{\arabic{section}.\arabic{subsection}.\arabic{equation}}
\makeatletter \@addtoreset{figure}{section}
\def\thefigure{\thesection.\@arabic\c@figure}
\def\fps@figure{h, t}
\@addtoreset{table}{bsection}
\def\thetable{\thesection.\@arabic\c@table}
\def\fps@table{h, t}
\@addtoreset{equation}{section}
\def\theequation{\thesection.\arabic{equation}}
\newtheorem{corollary}{Corollary}[section]
\newtheorem{definition}{Definition}[section]

\newtheorem{proposition}{Proposition}[section]

\newtheorem{examps}{Examples}[section]

\newtheorem{lemma}{Lemma}[section]
\newtheorem{remark}{Remark}[section]
\newtheorem{remarks}[remark]{Remarks}
\def\mod{\,\hbox{mod}\,}

\def\bx{\begin{example}}
\def\ex{\end{example}}
\def\bxs{\begin{examps}. \rm\begin{enumerate}}
\def\exs{\end{enumerate}\end{examps}}
\def\bd{\begin{definition}}
\def\ed{\end{definition}}
\def\bp{\begin{proposition}\rm}
\def\ep{\end{proposition}}
\def\bc{\begin{corollary}}
\def\ec{\end{corollary}}
\def\bl{\begin{lemma}\em}
\def\el{\end{lemma}}
\def\be{\begin{equation}}
\def\ee{\end{equation}}
\def\br{\begin{remark}\rm\small}
\def\er{\end{remark}}
\def\brs{\begin{remarks}.\\ \rm\
\begin{enumerate}}
\def\ers{\end{enumerate}\end{remarks}}
\def\bea{\begin{eqnarray}}
\def\eea{\end{eqnarray}}

\def\ra{{\rightarrow}}

\def\Tr{\mathrm {Tr}}
\def\tr{\mathrm {tr}}
\def\det{\mathrm {det}}
\def\sgn{\mathrm {sgn}}

\def\diag{\mathrm {diag}}

\def\&{&{\hskip -20pt}}

\def\AA{{\mathcal A}}

\def\VV{{\mathcal V}}

\def\Ib{{\mathbf I}}

\def\Nb{{\mathbf N}}

\def\Ub{{\mathbf U}}

\def\Zb{{\mathbf Z}}

\def\Nbb{{\mathbb N}}

\date{}
\begin{document}
\baselineskip 16pt
\begin{flushright}
CRM-3195 (2005)
\end{flushright}
\medskip
\begin{center}
\begin{Large}\fontfamily{cmss}
\fontsize{17pt}{27pt} \selectfont \textbf{Fermionic construction
of partition functions for two-matrix models and perturbative Schur
function expansions}\footnote{Work of (J.H.) supported in part by the
Natural Sciences and Engineering Research Council of Canada
(NSERC) and the Fonds FCAR du Qu\'ebec; that of (A.O.)
by the Russian Academy of Science program  ``Mathematical Methods in Nonlinear Dynamics" and  RFBR grant No 05-01-00498.}
\end{Large}\\
\bigskip
\begin{large}  {J. Harnad}$^{\dagger \ddagger}$\footnote{harnad@crm.umontreal.ca}
 and {A. Yu. Orlov}$^{\star}$\footnote{orlovs@wave.sio.rssi.ru}
\end{large}
\\
\bigskip
\begin{small}
$^{\dagger}$ {\em Centre de recherches math\'ematiques,
Universit\'e de Montr\'eal\\ C.~P.~6128, succ. centre ville, Montr\'eal,
Qu\'ebec, Canada H3C 3J7} \\
\smallskip
$^{\ddagger}$ {\em Department of Mathematics and
Statistics, Concordia University\\ 7141 Sherbrooke W., Montr\'eal, Qu\'ebec,
Canada H4B 1R6} \\
\smallskip
$^{\star}$ {\em Nonlinear Wave Processes Laboratory, \\
Oceanology Institute, 36 Nakhimovskii Prospect\\
Moscow 117851, Russia } \\
\end{small}
\end{center}
\bigskip
\bigskip

\begin{center}{\bf Abstract}
\end{center}
\smallskip

\begin{small}
A new  representation of the $2N$ fold integrals appearing  in various two-matrix models  that admit reductions to integrals over their eigenvalues is given in terms of vacuum state expectation values of operator products formed from two-component free fermions. This is used to derive the perturbation series for these integrals under deformations induced by exponential weight factors in the measure, expressed as  double and quadruple Schur function expansions,  generalizing results obtained earlier for certain two-matrix models. Links with the coupled two-component KP hierarchy and  the two-component Toda lattice hierarchy are also derived.
\end{small}
\bigskip

\section{Introduction}

Let $d\mu(x,y)$ be a measure (in general, complex), supported either on a
finite set of products of curves in the complex $x$ and $y$
planes or, alternatively, on a domain in the complex $z$ plane,
with the identifications $(x=z, y=\bar{z})$.
Consider the following  $2N$-fold integral
\begin{equation}\label{Z_N}
{ Z}_N=\int d\mu(x_1,y_1)
 \dots \int d\mu(x_N,y_N)
 \Delta_N(x)\Delta_N(y),
\end{equation}
evaluated  over some finite linear combination of support domains, where
  \be \Delta_N(x) = \prod_{i < j}^N (x_i-x_j), \quad
\Delta_N(y) = \prod_{i < j}^N (y_i-y_j)
 \ee
are Vandermonde determinants.

Such integrals arise, in particular, as partition functions  and correlation functions for two-matrix models in those cases where one can reduce the integration over the matrix
ensemble, e.g., via the Itzykson-Zuber,  Harish-Chandra  identity \cite{IZ}, to integrals over  the eigenvalues $\{x_i, y_i\}_{i =1,\dots,N}$ of the   two matrices. Depending on the
specific choice of measure and support, these include  models of normal matrices
\cite{CZ} with spectrum supported on some open region of the complex  plane,  coupled pairs of random hermitian matrices  \cite{Mehta, IZ} or, more generally,
normal matrices with spectrum supported  on  curve segments in the complex plane
\cite{BEH1, BEH2},  and certain models of random unitary  matrices \cite{ZJ, ZJZ}.
  (See Appendix \ref{AppendA} for several  examples.)

From the viewpoint of perturbation theory, and also in order to apply methods from the theory of integrable systems to the study of such integrals, it is of interest to consider deformations of the measure of the general form
\be
\label{measure-defor-2cTL}
 d\mu(x,y|{\bf t},n,m,{\bar{\bf t}})
:= x^n y^m e^{V(x,{\bf t}^{(1)})+V(y,{\bf
t}^{(2)})+V(x^{-1},\bar{\bf t}^{(1)})+V(y^{-1},\bar{\bf t}^{(2)})
}d\mu(x,y),
\ee
 where
\be
\label{V-x-t}
 V(x,{\bf t}^{(\alpha)})=\sum_{k=1}^\infty t_k^{(\alpha)}  x^k,\quad
  V(x^{-1},\bar{\bf t}^{(\alpha)})=\sum_{k=1}^\infty \bar{t}_k^{(\alpha)} x^{-k},\quad
  \alpha=1,2.
\ee
The four infinite sequences of complex numbers ${\bf
t}^{(1)}=(t_1^{(1)},t_2^{(1)},\dots)$, ${\bf
t}^{(2)}=(t_1^{(2)},t_2^{(2)},\dots)$, ${\bar{\bf
t}}^{(1)}=(\bar{t}_1^{(1)},\bar{t}_2^{(1)},\dots)$, ${\bar{\bf
t}}^{(2)}=(\bar{t}_1^{(2)},\bar{t}_2^{(2)},\dots)$ and the integers
$n,m$ are viewed as independent deformation parameters. (Thus the quantities $(\bar{\bf t}^{(1)}, \bar{\bf t}^{(2)})$ are not, in general,  complex conjugates
of $({\bf t}^{(1)}$ , ${\bf t}^{(2)}$).) For brevity, we also use the notation ${\bf t}:=({\bf t}^{(1)}, {\bf t}^{(2)})$, \hbox{$ {\bar{\bf t}}:=(\bar{\bf t}^{(1)}, \bar{\bf t}^{(2)})$}. The correspondingly deformed $Z_N$ will be denoted
$Z_N({\bf t},n,m,\bar{\bf t})$.

In the present work we represent the integral (\ref{Z_N}) and its deformations as  vacuum state expectation values  (VEV) using two-component fermions,
and use this to derive perturbation expansions for $Z_N({\bf t},n,m,\bar{\bf t})$
as series in products of two or four Schur functions, with the four sets of continuous deformation parameters $\{ {\bf t}^{(1)}, {\bf t}^{(2)}, \bar{\bf t}^{(1)},\bar {\bf t}^{(2)}\}$
as their arguments. Such character expansions have been studied for various special cases of one and two matrix models by a number of
   authors  \cite{KSW1, KSW, HO1, HO2, OS, ZJ, ZJZ, AvM} using a variety methods.
   The fermionic  approach presented here is based on the original
   constructions of $\tau$-functions for integrable hierarchies (refs. {\cite{DJKM,JM} and
   \cite{KdL}) as fermionic VEV's .

In section \ref{fermionic}, after presenting  the integral (\ref{Z_N}) in the form of a
fermionic vacuum expectation value, we show that the corresponding integral
$Z_N({\bf t},n,m,\bar{\bf t})$ for the deformed measure
(\ref{measure-defor-2cTL}) is a special case of a $\tau$-function for the coupled two-component KP hierarchy or, equivalently, a $\tau$-function for the two-component Toda lattice (TL) hierarchy \cite{DJKM, Takas, Take, UT}. (See (\ref{result2}) below.)

We also use this fermionic representation to derive the perturbation expansion for $Z_N({\bf t},n,m,\bar{\bf t})$ as weighted power series in the deformation
parameters, written in the following alternative forms as  sums
over products of Schur functions
\bea
\label{ssss}
Z_N({\bf t},n,m,\bar{\bf
t})&\&=\sum_{\lambda,\mu,\nu,\eta}I_{\lambda \mu \nu \eta}(N,n,m)s_\lambda({\bf
t}^{(1)})s_\mu({\bf t}^{(2)})s_\nu(\bar{\bf
t}^{(1)})s_\eta(\bar{\bf t}^{(2)}) \\
 \label{++'}
&\& =  N!\sum_{\lambda,\mu}
g_{\lambda \mu }^{++}(N,n,m,\bar{\bf t}^{(1)},\bar{\bf t}^{(2)}
)s_\lambda({\bf t}^{(1)})s_\mu({\bf t}^{(2)})
\\
&\&  =  N!\sum_{\lambda,\mu} g_{\lambda \mu }^{--}(N,n,m,{\bf
t}^{(1)},{\bf t}^{(2)})s_\lambda(\bar{\bf t}^{(1)})s_\mu(\bar{\bf
t}^{(2)}) \label{--'}
\\
&\& =   N!\sum_{\lambda,\mu} g_{\lambda \mu }^{+-}(N,n,m,\bar{\bf
t}^{(1)},{\bf t}^{(2)})s_\lambda({\bf t}^{(1)})s_\mu(\bar{\bf
t}^{(2)}) \label{+-'}
 \\
&\& =   N!\sum_{\lambda,\mu } g_{\lambda \mu }^{-+}(N,n,m,{\bf
t}^{(1)},\bar{\bf t}^{(2)})s_\lambda(\bar{\bf t}^{(1)})s_\mu({\bf
t}^{(2)}), \label{-+'}
  \eea
  where $s_\lambda(\bf t)$ is the Schur function  corresponding to a partition
  $\lambda:=(\lambda_1, \dots \lambda_N)$ of length $N=\ell(\lambda)$
   (see \cite{Mac}, or Appendix \ref{AppendB} ) and the sums are over all quadruples $(\lambda, \mu, \nu, \eta)$ or pairs $(\lambda, \mu)$ of such partitions. The coefficients $I_{\lambda \mu \nu \eta}(N,n,m)$,
   $g_{\lambda \mu }^{++}(N,n,m,\bar{\bf t}^{(1)},\bar{\bf t}^{(2)})$,
 $ g_{\lambda \mu }^{--}(N,n,m,{\bf t}^{(1)},{\bf t}^{(2)})$,
$g_{\lambda \mu }^{+-}(N,n,m,\bar{\bf t}^{(1)},{\bf t}^{(2)})$ and
$ g_{\lambda \mu }^{-+}(N,n,m,{\bf t}^{(1)},\bar{\bf t}^{(2)})$
 are expressed as determinants of $N \times N$  submatrices of the matrix of
 bimoments
   \be\label{bare-bi-moments}
B_{ik}=\int x^iy^{k}d\mu(x,y), \quad i,k \in {\bf Z}
 \ee
 and its deformations (see eqs.~(\ref{g++Nnm})-(\ref{bimoments}) and (\ref{Ilmne}) below).

For  normal matrix models,  formula (\ref{++'}) was presented
in \cite{OS}. The case of diagonal coefficients
$g_{\lambda \mu }=\delta_{\lambda \mu }r_\lambda$ was earlier
considered in \cite{HO1, HO2} in the context of models
of pairs of hermitian matrices with Itzykson-Zuber coupling \cite{IZ} and of normal matrices with axially symmetric interactions \cite{CZ}. In the present work, we
 derive  formulae (\ref{ssss})-(\ref{-+'}) by two different methods. First, in Section \ref{fermionic}  via the two component fermionic calculus and then, in
 Section \ref{direct},  by a direct calculation, as in \cite{HO1,HO2, OS}, based on a standard determinant integral identity (the Andr\'eief formula (\ref{detidentity})),
  combined with the Cauchy-Littlewood identity (\ref{Cauchy}).
  We also  present some of the earlier results in a
more complete way; in particular, the expansions
(\ref{++'})-(\ref{-+'}) are given for  the case of coupled pairs
of hermitian matrices.   The fermionic representation to be
used here however differs from the ones appearing previously
in the context of matrix models in refs.~\cite{ZKMMO, HO1, HO2}, where
the links with the one component KP (and TL) hierarchy were
developed. It is also different from the fermionic approach of ref.~\cite{CZ},
where two-dimensional fermions were used in the study of
normal matrix models.

\br One can obtain the $N$-fold integrals arising in one-matrix
models as specializations of the above, either by choosing the
undeformed measure to contain a factor proportional to the Dirac
delta function $\delta(x-y)$, in which case the matrix of
bimoments becomes a Hankel matrix, or to $\delta(x-{1 \over y})$, in
which case it becomes a Toeplitz matrix. We shall not consider
these specializations here.
\er

\subsection{Free fermions \label{freefermi}}

 Let $\AA$ be the complex Clifford algebra over $ \mathbb{C}$ generated
by \emph{charged free fermions} \hbox{$\{f_i$, ${\bar f}_i\}_{i\in
{\bf Z}}$},  satisfying the anticommutation relations
\begin{equation}\label{fermions}
[f_i,f_j]_+=[{\bar f}_i,{\bar f}_j]_+=0,\quad [f_i,{\bar
f}_j]_+=\delta_{ij}.
\end{equation}
Any element of the linear part \be W:=\left(\oplus_{m \in
\Zb}\mathbb{ C}f_m\right)\oplus \left(\oplus_{m\in
\Zb}\mathbb{ C}{\bar f}_m\right)
 \ee will be referred to as a {\em free fermion}. We also introduce the
 fermionic free fields
\be
\label{fermions-fourier}
    f(x):=\sum_{k\in\Zb}f_kx^k,\quad
    {\bar f}(y):=\sum_{k\in\Zb}{\bar f}_ky^{-k-1},
\ee
which may be viewed as generating functions for the $f_j, \bar{f}_j$'s.

This Clifford algebra has a standard Fock
space representation defined as follows.
Define the complementary, totally null (with respect to the
underlying quadratic form) and mutually dual subspaces
 \be W_{an}:=\left(\oplus_{m<0}\mathbb{
C}f_m\right)\oplus \left(\oplus_{m\ge 0}\mathbb{ C}{\bar
f}_m\right), \qquad  W_{cr}:=\left(\oplus_{m\ge
0}\mathbb{C}f_m\right)\oplus \left(\oplus_{m< 0}\mathbb{ C}{\bar
f}_m\right),
\ee
and consider the left  and right $\AA$-modules
\be
 F:=\AA/\AA W_{an}, \qquad {\bar F}:=W_{cr}\AA{\backslash}\AA.
\ee
These are cyclic $\AA$-modules generated by the vectors
 \be
|0\rangle= 1\  \mod \  \AA  W_{an}, \qquad  \langle 0|= 1 \  \mod
\ W_{cr}\AA ,
\ee
respectively, with the properties
\bea
\label{vak}
f_m |0\rangle=0 \qquad (m<0),\qquad {\bar f}_m|0\rangle =0 \qquad
(m \ge 0) , \cr
 \langle 0|f_m=0 \qquad (m\ge 0),\qquad \langle
0|{\bar f}_m=0 \qquad (m<0) .
\eea
The {\em Fock spaces} $F$ and ${\bar F}$ are mutually dual, with
the hermitian pairing defined via the linear form $\langle 0| |0 \rangle$ on
$\AA$ called the {\em vacuum expectation value}.  This is determined by
\bea
\label{psipsi*vac}
\langle 0|1|0 \rangle&\&=1;\quad \langle 0|f_m{\bar f}_m
|0\rangle=1,\quad m<0; \quad  \langle 0|{\bar f}_mf_m
|0\rangle=1,\quad m\ge 0 ,\\
\label{end}
 \langle 0| f_n
 |0\rangle&\&=\langle 0|{\bar f}_n
 |0\rangle=\langle 0|f_mf_n |0\rangle=\langle 0|{\bar f}_m{\bar f}_n
 |0\rangle=0;
 \quad \langle 0|f_m{\bar f}_n|0\rangle=0, \quad m\ne n,\cr
 &\&
\eea
together with the Wick theorem which implies, for any finite set of elements
$\{w_k \in W\}$,
\bea \label{Wick}
\langle 0|w_1 \cdots w_{2n+1}|0 \rangle &\&=0,\cr
 \langle 0|w_1
\cdots w_{2n} |0\rangle &\&=\sum_{\sigma \in S_{2n}} sgn\sigma \langle
0|w_{\sigma(1)}w_{\sigma(2)}|0\rangle \cdots \langle 0|
w_{\sigma(2n-1)}w_{\sigma(2n)} |0\rangle .
\eea
Here  $\sigma$ runs over permutations for which
$\sigma(1)<\sigma(2),\dots , \sigma(2n-1)<\sigma(2n)$ and
$\sigma(1)<\sigma(3)<\cdots <\sigma(2n-1)$.

Now let  $\{w_i\}_{ i=1,\dots,N}$, be linear combinations of the $f_j$'s only,
$j\in\Zb$, and  $\{{\bar w}_i\}_{ i=1,\dots,N}$
linear combinations of the ${\bar f}_j$'s, $j \in\Zb$. Then(\ref{Wick})
implies
\begin{equation}\label{Wick-det}
\langle 0|w_1\cdots w_{N}{\bar w}_N \cdots {\bar w}_1 |0\rangle
=\det\; (\langle 0| w_i{\bar w}_j|0\rangle)\ |_{i,j=1,\dots,N}
\end{equation}

Following refs.~\cite{DJKM},\cite{JM},  for all $ N\in \Zb$, we also  introduce the states
 \be \label{1-vacuum}
  \langle  N|:=\langle 0|C_{N}
 \ee
where
 \bea
\label{1-vacuum'} C_{N}&\&:={\bar f}_0\cdots {\bar
f}^{(\alpha)}_{N-1}
 \quad {\rm if }\ N>0 \\
C_{N}&\&:={ f}_{-1}\cdots { f}_{N}
\quad {\rm if}\ N<0  \\
 C_{N}&\&:=1 \quad {\rm if}\ N=0
 \eea
and
 \be \label{1-vacuum-r}
    |N \rangle:={\bar C}_{N}|0\rangle
 \ee
where
 \bea
\label{1-vacuum'-r} {\bar C}_{N}&\&:=f_{N-1}\cdots
f_0 \quad {\rm if }\ N>0 \\
{\bar C}_{N}&\&:={\bar f}_{N}\cdots {\bar f}_{-1}
\quad {\rm if}\ N<0  \\
 {\bar C}_{N}&\&:=1 \quad {\rm if}\ N=0
 \eea
The states   (\ref{1-vacuum}) and (\ref{1-vacuum-r}) are referred to as the
left and right charged vacuum vectors, respectively, with charge $N$.

In what follows we use the notational convention
\begin{equation}\label{Delta(N)}
\Delta_N(x)=\det \; (x_i^{N-k})|_{i,k=1,\dots,N}\ (N>0),\quad
\Delta_0(x)=1,\quad \Delta_N(x)=0\ (N<0).
\end{equation}
From the relations
 \be
 \langle 0|  {\bar f}_{N-k} f(x_{i})|0\rangle
 =x_i^{N-k},\quad  \langle 0|
 { f}_{-N+k-1}  {\bar f}(y_{i})|0\rangle =y_i^{N-k},\quad k=1,2,\dots ,
 \ee
 and  (\ref{Wick-det}), it follows that
\bea\label{Delta-N-left} \langle N|f(x_1)\cdots
f(x_n)|0\rangle  &\&=\delta_{n,N}\Delta_N(x),\quad N\in \Zb,\\
\label{Delta-N-right} \langle -N|\bar{f}(y_1)\cdots
\bar{f}(y_n)|0\rangle&\&=\delta_{n,N}\Delta_N(y),\quad N\in
\Zb. \eea

\subsection{Two-component fermions \label{2compfermi} }

The $2$-component fermion formalism is obtained by
relabelling  the above as follows.
 \bea
  \label{2-fermions}
f_n^{(\alpha)}&\&:=f_{2n+\alpha-1}\ ,\qquad \qquad {\bar
f}_n^{(\alpha)}:={\bar f}_{2n+\alpha-1}\ ,
\\
 \label{2-fermions-z}
f^{(\alpha)}(z)&\&:=\sum_{k=-\infty}^{+\infty}z^kf_{k}^{(\alpha)}\
,\quad {\bar
f}^{(\alpha)}(z):=\sum_{k=-\infty}^{+\infty}z^{-k-1}{\bar
f}_{k}^{(\alpha)}\ ,
 \eea
where $\alpha=1,2$. Then (\ref{fermions}) is equivalent to
\begin{equation}\label{2-fermions-antic}
[f_n^{(\alpha)},f_m^{(\beta)}]_+=[{\bar f}_n^{(\alpha)},{\bar
f}_m^{(\beta)}]_+=0,\qquad [f_n^{(\alpha)},{\bar
f}_m^{(\beta)}]_+=\delta_{\alpha,\beta}\delta_{nm}.
\end{equation}

We denote the right and left vacuum vectors respectively as
\be \label{2-vacuum-def}
|0,0\rangle:=|0\rangle ,\quad \langle 0,0|:=\langle 0| .
\ee
Relations (\ref{vak}) then become, for $\alpha=1,2$,
\begin{eqnarray}\label{2-vak-r}
f_m^{(\alpha)} |0,0\rangle=0 \qquad (m<0),\qquad {\bar
f}_m^{(\alpha)}|0,0\rangle =0 \qquad (m \ge 0) , \\
\label{2-vak-l} \langle 0,0|f_m^{(\alpha)}=0 \qquad (m\ge
0),\qquad \langle 0,0|{\bar f}_m^{(\alpha)}=0 \qquad (m<0) .
\end{eqnarray}

As in \cite{DJKM},\cite{JM}, we also introduce the states
 \be \label{2-vacuum}
  \langle  n^{(1)},n^{(2)}|:=\langle 0,0|C_{n^{(1)}}C_{n^{(2)}},
 \ee
where
\bea
\label{2-vacuum'} C_{n^{(\alpha)}}&\&:={\bar f}^{(\alpha)}_0\cdots
{\bar f}^{(\alpha)}_{n^{(\alpha)}-1}
 \quad {\rm if }\ n^{(\alpha)}>0 \\
C_{n^{(\alpha)}}&\&:={ f}^{(\alpha)}_{-1}\cdots {
f}^{(\alpha)}_{n^{(\alpha)}}
 \qquad {\rm if}\ n^{(\alpha)}<0  \\
 C_{n^{(\alpha)}}&\&:=1 {\hskip 79 pt} {\rm if}\ n^{(\alpha)}=0
 \eea
and
 \be \label{2-vacuum-r}
    |n^{(1)},n^{(2)}\rangle:={\bar C}_{n^{(2)}}{\bar C}_{n^{(1)}}|0,0\rangle
 \ee
where
 \bea
\label{2-vacuum'-r} {\bar
C}_{n^{(\alpha)}}&\&:=f^{(\alpha)}_{n^{(\alpha)}-1}\cdots
f^{(\alpha)}_0 \quad {\rm if }\ n^{(\alpha)}>0 \\
{\bar C}_{n^{(\alpha)}}&\&:={\bar
f}^{(\alpha)}_{n^{(\alpha)}}\cdots {\bar f}^{(\alpha)}_{-1}
\qquad {\rm if}\ n^{(\alpha)}<0  \\
 {\bar C}_{n^{(\alpha)}}&\&:=1  {\hskip 79 pt}  {\rm if}\ n^{(\alpha)}=0
 \eea
The states (\ref{2-vacuum}) and (\ref{2-vacuum-r}) will be
referred to, respectively, as left and right charged vacuum
vectors with charges $(n^{(1)},n^{(2)})$.

It is  easily verified that
\begin{eqnarray}\label{2-vak-ch-1-r}
f_m^{(1)} |n,*\rangle=0 \qquad (m<n),\qquad {\bar
f}_m^{(1)}|n,*\rangle =0 \qquad (m \ge n) , \\
\label{2-vak-ch-1-l} \langle n,*|f_m^{(1)}=0 \qquad (m\ge
n),\qquad \langle n,*|{\bar f}_m^{(1)}=0 \qquad (m<n) ,
\end{eqnarray}
and similarly
\begin{eqnarray}\label{2-vak-ch-2-r}
f_m^{(2)} |*,n\rangle=0 \qquad (m<n),\qquad {\bar
f}_m^{(2)}|*,n\rangle =0 \qquad (m \ge n) , \\
\label{2-vak-ch-2-l} \langle *,n|f_m^{(2)}=0 \qquad (m\ge
n),\qquad \langle *,n|{\bar f}_m^{(2)}=0 \qquad (m<n) .
\end{eqnarray}

\br
\label{2Wick}
 In subsequent calculations we use  Wick's theorem in the form
(\ref{Wick-det}). In the two component setting we use formula (\ref{Wick-det}) separately
for each component. To calculate the vacuum expectation
value of an operator $O$, first express it in the form
 \be\label{O-comp-decomp}
O=\sum_i O^{(1)}_i O^{(2)}_i.
 \ee
Then
 \be
 \label{Wick-comp-decomp}
\langle 0,0|O|0,0\rangle=\sum_i \langle 0,0|O^{(1)}_i O^{(2)}_i
|0,0\rangle=\sum_i \langle 0,0|O^{(1)}_i|0,0\rangle \langle
0,0|O^{(2)}_i |0,0\rangle
 \ee
where Wick's theorem in the form (\ref{Wick-det}) may be applied to each
factor $\langle 0,0|O^{(\alpha)}_i|0,0\rangle$.
\er

As a first application, note that Wick's theorem and (\ref{Delta-N-left})-(\ref{Delta-N-right}) imply that
 \be\label{D-D}
  \langle N,-N| \prod_{i=1}^k  f^{(1)}(x_i){\bar
f}^{(2)}(y_i)|0,0\rangle= (-1)^{\frac 12
N(N+1)}\delta_{k,N}\Delta_N(x)\Delta_N(y),
 \ee
where $\Delta_N(x)$ is defined in (\ref{Delta(N)}).
It  then easily follows that
 \be\label{vacuum-ff}
  \langle N+n,-N-m| \prod_{i=1}^k  f^{(1)}(x_i){\bar
f}^{(2)}(y_i)|n,-m\rangle= (-1)^{\frac 12
N(N+1)}\delta_{k,N}\Delta_N(x)\Delta_N(y)\prod_{i=1}^N
x_i^n(-y_i)^m.
 \ee


\section{Fermionic representation for $Z_N({\bf t},n,m,\bar{\bf t})$
and double Schur function expansions \label{fermionic}}


\subsection{Fermionic representation for $Z_N$}

Let
\begin{equation}\label{A-for-2MM}
A:=\int f^{(1)}(x){\bar f}^{(2)}(y)d\mu(x,y).
\end{equation}
Then the following gives a fermionic VEV representation of $Z_N$
 \be\label{Z=ev}
Z_N=(-1)^{\frac 12 N(N+1)}\langle N,-N| A^N|0,0\rangle
 \ee
To see this, just use (\ref{D-D}) to write
\bea
\langle N,-N| A^N|0,0\rangle
&\& = \langle N,-N| \prod_{i=1}^N \int f^{(1)}(x_i){\bar
f}^{(2)}(y_i)d\mu(x_i,y_i)|0,0\rangle
\\
&\& = (-1)^{\frac 12
N(N+1)} \int \dots \int \Delta_N(x)\Delta_N(y) \prod_{i=1}^N
d\mu(x_i,y_i),
\eea
where the last equality follows from  (\ref{D-D}).
Now introduce
  \be
   \label{el-g}
 g:=e^{A}:=\sum_{k=0}^\infty \frac{A^k}{k!}.
  \ee
  Since
  \be
  \langle N,-N| A^k|0,0\rangle = 0 \qquad {\rm if} \quad k\neq N
  \ee
we may equivalently express (\ref{Z=ev}) as
 \be\label{result1}
Z_N= (-1)^{\frac 12 N(N+1)}N!\langle N,-N| g|0,0\rangle ,\quad
N\ge 0
 \ee
By (\ref{D-D}) we also obtain
 \be\label{N<0}
\langle N,-N|e^{A} |0,0\rangle=0,\quad N<0 .
 \ee

Next we  show that for the deformed measure
(\ref{measure-defor-2cTL}) the expression (\ref{result1}) (or, equivalently (\ref{Z=ev}))
determines a $\tau$-function in the sense of integrable hierarchies.


\subsection{ $Z_N({\bf t},n,m,\bar{\bf t})$ as a $\tau$-function \cite{DJKM} }

We begin by recalling briefly the important notion of
$\tau$-functions as  introduced by the Sato school. (See \cite{DJKM, JM}.)
First, we define two infinite linear families  of operators
$H({\bf t})$ $\bar{H}(\bar{\bf t})$ by
 \be\
\label{H(t)}
H({\bf t}):= \sum_{k=1}^\infty H_k^{(1)}
t_k^{(1)}-\sum_{k=1}^\infty H_k^{(2)} t_k^{(2)},\qquad
\bar{H}(\bar{\bf t}):= \sum_{k=1}^\infty H_{-k}^{(2)}
\bar{t}_k^{(2)}-\sum_{k=1}^\infty H_{-k}^{(1)} \bar{t}_k^{(1)}
 \ee
 where the  ``commuting Hamiltonians'' $H_{k}^{(\alpha)}$ are bilinear combinations of
 fermion components of the form
 \be
\label{ham-2} H_{k}^{(\alpha)}:=\sum_{n=-\infty}^{+\infty}
f_n^{(\alpha)}{\bar f}_{n+k}^{(\alpha)} \ ,\quad k\neq 0, \quad \alpha=1,2.
 \ee

Now, let
\be
\label{A-gen} g:=
exp \sum_{\alpha,\beta=1,2}\int \; :f^{(\alpha)}(x){\bar
f}^{(\beta)}(y):d\mu_{\alpha \beta}(x,y),
\ee
where $\{d\mu_{\alpha \beta}\}$ is some $2 \times 2$ matrix of measures, and
\be
: f^\alpha \bar{f}^\beta:=  f^\alpha \bar{f}^\beta - <0,0|  f^\alpha \bar{f}^\beta |0,0>.
\ee
Then the expectation value
 \be
  \label{2-comp-TL-tau}
 \tau_N({\bf t},n,m,\bar{\bf t})=\langle N+n,-N-m|e^{H({\bf t})}
 g e^{\bar{H}(\bar{\bf t})} |n,-m\rangle ,\quad
 N,n,m\in\Zb,
 \ee
if it exists, is called the $\tau$-function of the two-component Toda Lattice (TL)
hierarchy or, equivalently, the coupled two-component KP hierarchy.
(See \cite{DJKM, UT}.)

 The sets of parameters ${\bf t},\bar{\bf t}$ and also
 the integers $N$, $n$ and $m$ are called ``higher times'' of the two-component TL hierarchy. The parameters ${\bf t}=({\bf t}^{(1)},{\bf t}^{(2)})$ are higher
times of the two-component KP hierarchy, and the second set $\bar{\bf t}=(\bar{\bf
t}^{(1)}, \bar{\bf t}^{(2)})$ is also a set of higher times, for  a different two-component KP hierarchy. Together, they may be referred to as the coupled (two-component) KP hierarchy. The $\tau$-function (\ref{2-comp-TL-tau}) solves an infinite number of bilinear (Hirota) equations. (Further details may be found in refs.~\cite{DJKM, JM, KdL}. We only note here that each equation contains a certain number of derivatives of
the $\tau$-function with respect to the variables $t_k^{(\alpha)}$
and $\bar{t}_k^{(\alpha)}$,  $\alpha=1,2$,   $k=1,2,\dots$, and the equations
may also relate $\tau$-functions with different values of the variables $N,n,m$.)

To relate $Z_N({\bf t},n,m,\bar{\bf t})$ to the $\tau$-function in the above sense, we choose the measures $d\mu_{11}$, $d\mu_{22}$ and
$d\mu_{21}$ to all vanish, and $d\mu_{12}:= d\mu$.
Thus $g=e^A$ is of the form (\ref{el-g}) where $A$ is defined by (\ref{A-for-2MM}).
(In this case   $:A:$ just coincides with $A$.)
We now will prove that, for $N\ge 0$, the resulting $\tau$-function,
 up to a simple explicit multiplicative factor,  is equal to $Z_N({\bf t},n,m,\bar{\bf t})$,
 where the measure $d\mu$ is the deformed one in (\ref{measure-defor-2cTL}):
  \bea
 \tau_N({\bf t},n,m,\bar{\bf t})&\&:=\langle N+n,-N-m|e^{H({\bf t})}
 e^A e^{\bar{H}(\bar{\bf t})} |n,-m\rangle \\
&\&= {1\over N!}(-1)^{{1\over 2} N(N+1)+mN}c({\bf t},\bar{\bf t})
Z_N({\bf t},n,m,\bar{\bf t})
  \label{result2} \\
&\&={1\over N!}(-1)^{{1\over 2} N(N+1)+mN}c({\bf t},\bar{\bf t})\int\dots\int\Delta_N(x)\Delta_N(y) \prod_{k=1}^N
d\mu(x_k,y_k|{\bf t},n,m,\bar{\bf t}),\cr
&\&
\eea
 where
\be
 c({\bf t},\bar{\bf t}):=e^{-\sum_{\alpha=1}^2
\sum_{k=1}^\infty k t_k^{(\alpha)} \bar{t}_k^{(\alpha)}}.
 \ee

\noindent
{\bf Proof:}
Eqns.~(\ref{eq:time evolution}) and (\ref{eq:time evolution bar})) of Appendix \ref{AppendB} imply that
 \bea
 e^{H({\bf t})}  f^{(1)}(x){\bar f}^{(2)}(y) e^{-H({\bf t})} &\&=
 e^{V(x,{\bf t}^{(1)})+ V(y,{\bf t}^{(2)})}f^{(1)}(x){\bar
f}^{(2)}(y), \\
 e^{-\bar{H}(\bar{\bf t})}  f^{(1)}(x){\bar f}^{(2)}(y) e^{\bar{H}(\bar{\bf t})} &\&=
 e^{V(x^{-1},\bar{\bf t}^{(1)})+ V(y^{-1},\bar{\bf t}^{(2)})}f^{(1)}(x){\bar
f}^{(2)}(y).
 \eea
Using definition (\ref{ham-2}) one can check that
 \be
e^{H({\bf t})}|n,m\rangle =|n,m\rangle,\quad \langle
n,m|e^{H(\bar{\bf t})} =\langle n,m|,\quad n,m\in\Zb
 \ee
and, from  the Heisenberg algebra relations
\be
[H_k^{(\alpha)},H_l^{(\beta)}]=k\delta_{\alpha,\beta}\delta_{k,-l},\quad
k,l =\pm 1,\pm 2,\dots,\quad \alpha,\beta=1,2,
\ee
it follows that
\be
e^{H({\bf t})}e^{H(\bar{\bf t})}= e^{-\sum_{\alpha=1}^2
\sum_{k=1}^\infty k t_k^{(\alpha)} \bar{t}_k^{(\alpha)}}
e^{H(\bar{\bf t})}e^{H({\bf t})}.
\ee
Combining these relations, we have
 \be\label{evt=ev}
\langle N+n,-N-m|e^{H({\bf t})}
 g e^{\bar{H}(\bar{\bf t})} |n,-m\rangle
 =e^{-\sum_{\alpha=1}^2 \sum_{k=1}^\infty k
t_k^{(\alpha)}\bar{t}_k^{(\alpha)}}\langle N +n,-N-m|e^{A({\bf
t},0,0,\bar{\bf t})}
  |n,-m\rangle ,
 \ee
 where
\begin{equation}\label{A-for-2MM-deformed}
A({\bf t},n,m,\bar{\bf t}):=\int f^{(1)}(x){\bar
f}^{(2)}(y)d\mu(x,y|{\bf t},n,m,\bar{\bf t}).
\end{equation}
Finally, to remove the numbers $n$ and $m$ from the vacuum states we
use (\ref{vacuum-ff}), giving
\be\label{evt=ev}
\langle N+n,-N-m|e^{H({\bf t})}
 g e^{\bar{H}(\bar{\bf t})} |n,-m\rangle
 =e^{-\sum_{\alpha=1}^2 \sum_{k=1}^\infty k
t_k^{(\alpha)}\bar{t}_k^{(\alpha)}}(-1)^{mN}\langle N
,-N|e^{A({\bf t},n,m,\bar{\bf t})}
  |0,0\rangle
 \ee
 Now (\ref{result2}) follows from (\ref{evt=ev})
 in the same way that (\ref{result1}) was shown equivalent to (\ref{Z=ev}).

\subsection{Perturbation series in the variables
${\bf t}^{(\alpha)},\bar{\bf t}^{(\alpha)}$ }

In this section we derive the expansions
  \bea \label{++}
Z_N({\bf t},n,m,\bar{\bf t}) &\& = N!\sum_{\lambda,\mu \atop
\ell(\lambda), \ell(\mu)\le N} g_{\lambda \mu }^{++}(N,n,m,\bar{\bf
t}^{(1)},\bar{\bf t}^{(2)} )s_\lambda({\bf t}^{(1)})s_\mu({\bf
t}^{(2)})
\\
&\&  =  N!\sum_{\lambda,\mu \atop \ell(\lambda), \ell(\mu)\le
N} g_{\lambda \mu }^{--}(N,n,m,{\bf t}^{(1)},{\bf
t}^{(2)})s_\lambda(\bar{\bf t}^{(1)})s_\mu(\bar{\bf t}^{(2)})
\label{--}
\\
&\& =   N!\sum_{\lambda,\mu \atop \ell(\lambda), \ell(\mu)\le
N} g_{\lambda \mu }^{+-}(N,n,m,\bar{\bf t}^{(1)},{\bf
t}^{(2)})s_\lambda({\bf t}^{(1)})s_\mu(\bar{\bf t}^{(2)})
\label{+-}
 \\
&\& =   N!\sum_{\lambda,\mu \atop \ell(\lambda), \ell(\mu)\le
N} g_{\lambda \mu }^{-+}(N,n,m,{\bf t}^{(1)},\bar{\bf
t}^{(2)})s_\lambda(\bar{\bf t}^{(1)})s_\mu({\bf t}^{(2)}),
\label{-+}
  \eea
where the sums range over all pairs of partitions
$\lambda=(\lambda_1,\dots,\lambda_N)$,
$\mu=(\mu_1,\dots,\mu_N)$ of lengths $\le N$. For such partitions, we also define
the labels\begin{equation}\label{hh'}
h_i=\lambda_i-i+N,\quad h'_i=\mu_i-i+N.
\end{equation}
In terms of these quantities, define four $N\times N$ determinants
 \bea
\label{g++Nnm}
 &\& g_{\lambda \mu }^{++}(N,n,m,\bar{\bf t}^{(1)},\bar{\bf t}^{(2)}):=
 \det\; (B_{n+h_i,m+h_j'}(0,0,\bar{\bf t}^{(1)},\bar{\bf t}^{(2)}))|_{i,j=1,\dots , N} ,
 \\
 &\&  \label{g--Nnm}
 g_{\lambda \mu }^{--}(N,n,m,{\bf t}^{(1)},{\bf t}^{(2)}):=
 \det\; (B_{N+n-h_i-1,N+m+1-h_j'}({\bf t}^{(1)},{\bf t}^{(2)},0,0))|_{i,j=1,\dots , N} ,
   \\
 &\& \label{g+-Nnm}
   g_{\lambda \mu }^{+-}(N,n,m,{\bf t}^{(2)},\bar{\bf t}^{(1)}):=c\;
 \det\; (B_{n+h_i,N+m+1-h_j'}(0,{\bf t}^{(2)},\bar{\bf t}^{(1)},0))|_{i,j=1,\dots , N} ,
 \\
 &\&  \label{g-+Nnm}
    g_{\lambda \mu }^{-+}(N,n,m,{\bf t}^{(1)},\bar{\bf
    t}^{(2)}):=c\;
 \det\; (B_{N+n-h_i-1,m+h_j'}({\bf t}^{(1)},0,0,\bar{\bf t}^{(2)}))|_{i,j=1,\dots ,
 N},
 \eea
where $c:=(-1)^{\frac12 N(N-1)}$, and
 \be
\label{bimoments} B_{i,k}({\bf t}^{(1)},{\bf t}^{(2)},\bar{\bf
t}^{(1)},\bar{\bf t}^{(2)}):= \int x^iy^{k}e^{V(x,{\bf
t}^{(1)})+V(y,{\bf t}^{(2)})+V(x^{-1},\bar{\bf
t}^{(1)})+V(y^{-1},\bar{\bf t}^{(2)}) }d\mu(x,y),\quad
i,k\in\Nb,
 \ee
  is the matrix of deformed bimoments.

 The series (\ref{++})-(\ref{-+}) show that we actually have a ``quadruple'' system of
 one-component TL hierarchies, since each of the double Schur series
 (\ref{++})--(\ref{-+}) has the  form of a Takasaki  expansion \cite{Takas,Take} of a
one-component TL $\tau$-function, where the sets of higher time
 variables  are, respectively, $({\bf t}^{(1)},{\bf
t}^{(2)})$,$(\bar{\bf t}^{(1)},\bar{\bf t}^{(2)})$,$({\bf
t}^{(1)},\bar{\bf t}^{(2)})$ and $(\bar{\bf t}^{(1)},{\bf
t}^{(2)})$. One can also view $Z_N({\bf t},n,m,\bar{\bf t})$ as
determining a  quadruple system of KP  $\tau$-functions, in which
each set, ${\bf t}^{(1)}$, ${\bf t}^{(2)}$, $\bar{\bf t}^{(1)}$ and $\bar{\bf t}^{(2)}$ plays the
role of (one-component) KP higher times.

We begin by proving (\ref{++}). First consider the case $n=m=0$
and $\bar{\bf t}^{(1)}= \bar{\bf t}^{(2)}=0$, then
 \be\label{Z=ev-t}
Z_N({\bf t}, 0, 0, \bar{\bf 0})=(-1)^{\frac 12 N(N+1)}\langle
N,-N|e^{H({\bf t})} A^N |0,0\rangle.
 \ee
Rewrite
 \be
  A =\int f^{(1)}(x){\bar
f}^{(2)}(y)d\mu(x,y)
\ee
 in terms of the component
operators as
 \be\label{Afbarf}
 A  =\sum_{i,k\in\Zb}f^{(1)}_i{\bar
f}^{(2)}_{-k-1}B_{ik},
 \ee
 where
 \be\label{bare-bi-moments}
B_{ik}=\int x^iy^{k}d\mu(x,y)
 \ee
are the bimoments.

 Because of the anticommutation relations
  \be
[f_j^{(1)},{\bar f_k}^{(2)}]_+  =0, \quad \forall j, k \in {\bf Z}
\ee and \be
 f^{(1)}_{-k-1}|0,0\rangle
=0,\quad {\bar f}^{(2)}_{k}|0,0\rangle=0, \quad   k\ge 0,
 \ee
nothing is changed by making the substitution
\be
\label{A++}
A \to A_{++}:= \sum_{i,k\ge 0}f^{(1)}_i{\bar f}^{(2)}_{-k-1}B_{ik},
\ee
(``projection to the creative corner'')  when evaluating the $\tau$-function.
To compute
 \be
 \langle N,-N| e^{H^{(1)}({\bf t}^{(1)})-
H^{(2)}({\bf t}^{(2)})} (A_{++})^N|0,0\rangle, \label{Apower}
 \ee
 we use the relation
\begin{equation}
\langle N,-N| e^{H^{(1)}({\bf t}^{(1)})- H^{(2)}({\bf t}^{(2)})}
f^{(1)}_{h_1}{\bar f}^{(2)}_{-h'_1-1}\cdots f^{(1)}_{h_N} {\bar
f}^{(2)}_{-h'_N-1}|0,0\rangle=(-1)^{\frac 12
N(N+1)}s_\lambda({\bf t}^{(1)})s_\mu({\bf t}^{(2)}),
\label{schurfunctionproduct}
\end{equation}
where $\lambda=(\lambda_1,\dots,\lambda_N)$ and
$\mu=(\mu_1,\dots,\mu_N)$ are the partitions
related to the labels $\{h_i, h'_j\}$ by (\ref{hh'}),
which is proved in Appendix \ref{AppendB} using Wick's theorem.

Note that each term (\ref{schurfunctionproduct}), for any given pair of
decreasing sets of non-negative integers
\be
\label{decrease}
h_1>h_2 \cdots >h_N \ge 0, \qquad h'_1>h'_2 \cdots >h'_N \ge 0
\ee
occurs in  (\ref{Apower}) $(N!)^2$ times, multiplied by a product of the moments,
with the sign determined by the permutations of the ordering,
\be
(-1)^{\sgn(\sigma)} (-1)^{\sgn(\tilde{\sigma)}}B_{h_\sigma(1), \tilde{\sigma}(h'_1)} \cdots B_{h_\sigma(N), \tilde{\sigma}(h'_N)},
\ee
where $\sigma, \tilde{\sigma} \in S_N$ are the permutations of the indices.
 (The index sets consist of distinct elements, since all VEV's with repeated values of $h_i$ or $h'_i$ vanish, and re-ordering  distinct sets
to satisfy (\ref{decrease}) changes the terms by the sign of the permutation.)
  Substitution of  (\ref{schurfunctionproduct}) into  (\ref{Apower}) thus gives
 \be
 \langle N,-N| e^{H^{(1)}({\bf t}^{(1)})- H^{(2)}({\bf
t}^{(2)})} A^N|0,0\rangle=N!\sum_{\lambda,\mu \atop \ell(\lambda), \ell\mu)\le N}
g^{++}_{\lambda \mu }(N)s_\lambda({\bf t}^{(1)})s_\mu({\bf t}^{(2)}),
 \ee
where
 \be
\label{g-N}
 g^{++}_{\lambda \mu }(N)=\det\; (B_{h_i,h'_j}) ,
  \quad
i,j=1,\dots , N,
 \ee
for partitions with $ \ell (\lambda ),\ell (\mu ) \le N$. Finally, we restore the dependence on the remaining variables ($n$,
$m$, and higher times $(\bar{\bf t}^{(1)},\bar{\bf t}^{(2)})$)
by including it in the measure. We thus obtain
 \be
 \langle N+n,-N-m| e^{H({\bf t})} A^N(\bar{\bf t})|n,-m\rangle=
 N!\sum_{\lambda,\mu \atop \ell(\lambda), \ell(\mu)\le N}
g^{++}_{\lambda \mu }(N,n,m,\bar{\bf t})s_\lambda({\bf
t}^{(1)})s_\mu({\bf t}^{(2)}),
 \ee
where $A(\bar{\bf t})=A({\bf 0},0,0,\bar{\bf t})$ (see
(\ref{A-for-2MM-deformed})) and where
 \be g^{++}_{\lambda \mu }(N,n,m,\bar{\bf t}^{(1)},\bar{\bf t}^{(2)})
 =\det\; (B_{h_i+n,h'_j+m}(\bar{\bf t}^{(1)},\bar{\bf t}^{(2)}) ,
  \quad
i,j=1,\dots,N, \label{g-N-n-m}
 \ee
 if $ \ell (\lambda ),\ell (\mu ) \le N$.

Relations (\ref{--})--(\ref{-+}), are proved similarly.
First, we equate the variables $n$, $m$ and all irrelevant
higher times to zero.  (For the cases (\ref{--}),(\ref{+-}) and (\ref{-+}),
these  times are $({\bf t}^{(1)},{\bf t}^{(2)})$,
$(\bar{\bf t}^{(1)},{\bf t}^{(2)})$ and $({\bf t}^{(1)},\bar{\bf
t}^{(2)})$,  respectively.) Then, for the cases  (\ref{--}),(\ref{+-})
and (\ref{-+}), we consider
 \bea\label{A--} &\& A \to A_{--}:= \sum_{i,k\ge 0}f^{(1)}_{N-i-1}{\bar
f}^{(2)}_{k-N}B_{N-i-1,1-k+N} ,
\\
&\& \label{A+-} A \to A_{+-}:= \sum_{i\ge,k\ge
0}f^{(1)}_{N+i}{\bar f}^{(2)}_{k-N}B_{N+i,1-k+N} ,
\\
&\& \label{A-+} A \to A_{-+}:= \sum_{i\ge,k\ge
0}f^{(1)}_{N-i-1}{\bar f}^{(2)}_{-k-1-N}B_{N-i-1,k+N}
 \eea
  and evaluate
 \bea &\&  \langle N,-N| (A_{--})^Ne^{  H^{(2)} (\bar{\bf
t}^{(2)})- H^{(1)} (\bar{\bf t}^{(1)})}|0,0\rangle,
\label{Apower--}
 \\
  &\&  \langle N,-N| e^{H^{(1)}({\bf t}^{(1)})} (A_{+-})^Ne^{  H^{(2)} (\bar{\bf
t}^{(2)})}|0,0\rangle, \label{Apower+-}
 \\
  &\&  \langle N,-N| e^{-
H^{(2)}({\bf t}^{(2)})} (A_{-+})^Ne^{ - H^{(1)} (\bar{\bf
t}^{(1)})}|0,0\rangle \label{Apower-+}.
 \eea

For the evaluation of (\ref{Apower--}), (\ref{Apower+-}) and
(\ref{Apower-+}) we use the following formulae (see Appendix \ref{AppendB})
 \bea
  &\& \langle N,-N|
f^{(1)}_{N-h_1-1}{\bar f}^{(2)}_{h'_1-N}\cdots f^{(1)}_{N-h_N-1}
{\bar f}^{(2)}_{h'_N-N}e^{  H^{(2)} (\bar{\bf t}^{(2)})- H^{(1)}
(\bar{\bf t}^{(1)})}|0,0\rangle \label{schurfunctionproduct--}
\\
&\& \quad =(-1)^{\frac 12 N(N+1)}s_\lambda(\bar{\bf
t}^{(1)})s_\mu(\bar{\bf t}^{(2)}), \nonumber
 \\
&\& \langle N,-N| e^{H^{(1)}({\bf t}^{(1)})} f^{(1)}_{h_1}{\bar
f}^{(2)}_{h'_1-N}\cdots f^{(1)}_{h_N} {\bar f}^{(2)}_{h'_N-N}e^{
H^{(2)} (\bar{\bf t}^{(2)})}|0,0\rangle
\label{schurfunctionproduct+-}
\\
&\&  \quad =(-1)^N s_\lambda({\bf t}^{(1)})s_\mu(\bar{\bf t}^{(2)}), \nonumber
\\
&\& \langle N,-N| e^{- H^{(2)}({\bf t}^{(2)})}
f^{(1)}_{N-h_1-1}{\bar f}^{(2)}_{-h'_1-1}\cdots f^{(1)}_{N-h_N-1}
{\bar f}^{(1)}_{-h'_N-1}e^{  - H^{(1)} (\bar{\bf
t}^{(1)})}|0,0\rangle \label{schurfunctionproduct-+}
\\
&\&  \quad =(-1)^Ns_\lambda(\bar{\bf t}^{(1)})s_\mu({\bf t}^{(2)}), \nonumber
 \eea
where $\lambda=(\lambda_1,\dots,\lambda_N)$ and
$\mu=(\mu_1,\dots,\mu_N)$ are again  the partitions
related to the labels $\{h_i, h'_j\}$ by (\ref{hh'}).

Finally, we restore the dependence on the remaining variables (($n, m)$
and the appropriate higher times, which are respectively $({\bf
t}^{(1)},{\bf t}^{(2)})$, $(\bar{\bf t}^{(1)},{\bf t}^{(2)})$ and
$({\bf t}^{(1)},\bar{\bf t}^{(2)})$ by including these in the definition of
the measure, to obtain (\ref{--}),(\ref{+-}) and (\ref{-+}).


\subsection{Generalized $2$-matrix models with polynomial potentials
\label{generalized2matrix}}
In the case of the standard exponentially coupled hermitian two-matrix model \cite{IZ},
and a pair of polynomial {\it potentials} of degrees $p+1$, $q+1$, or the generalized case
 of pairs of normal matrices (2NMM) with spectra supported on some specified curve segments (see refs.\cite{BEH1, BEH2}), we may consider the unperturbed measure  as corresponding to the leading monomial potentials
 \be
d\mu(x,y):=e^{-\frac {u_{p+1}}{p+1}x^{p+1}-\frac
{u_{q+1}}{q+1}y^{q+1}+xy}dxdy ,
 \ee
\begin{equation}
\label{generalized2Mintegral}
\int (*) d\mu(x,y)=
\sum_{a=1}^{p+1}\sum_{b=1}^{q+1}\kappa_{ab}\int_{\gamma_a}\int_{\Gamma_b}
(*)e^{-\frac {u_{p+1}}{p+1}x^{p+1}-\frac
{v_{q+1}}{q+1}y^{q+1}+xy}dxdy ,
\end{equation}
with the integration contours $\{\gamma_a,{\Gamma}_b\}$ chosen in such a way
 that all the bimoment integrals are finite:
\begin{equation}\label{bimoments-BEH}
B_{jk}:=\int x^j y^k d\mu(x,y)=
\sum_{i=1}^{p+1}\sum_{j=1}^{q+1}\kappa_{ab}\int_{\gamma_a}\int_{\Gamma_b}
x^j y^k e^{-\frac {u_{p+1}}{p+1}x^{p+1}-\frac
{v_{q+1}}{q+1}y^{q+1}+xy}dxdy \le \infty.
\end{equation}

Using the notations of refs.~\cite{BEH1, BEH2}, the deformations of the measure may be chosen to be exponentials  of lower degree polynomials,  defined by parameters
$u_1,\dots,u_p$ and $v_1,\dots,v_q$,   as well as multiplicative monomials in $x$ and $y$ of degrees $n,m\in\Nbb$
 \be
d\mu(x,y|{\bf u},{\bf v},n,m):=(-1)^mx^ny^me^{-\sum_{i=1}^p\frac
{u_{i}}{i}x^{i}-\sum_{i=1}^q\frac {v_{i}}{i}y^{i}}e^{-\frac
{u_{p+1}}{p+1}x^{p+1}-\frac {u_{q+1}}{q+1}y^{q+1}+xy}dxdy.
 \ee

Formula (\ref{++}) is applicable in this case  if we put
$\bar{\bf t}^{(1)}=\bar{\bf t}^{(2)}=0$ and define  ${\bf t}^{(1)}$
and ${\bf t}^{(1)}$ as
\bea
{\bf t}^{(1)}&\&={\bf t}_{\bf
u}:=(-u_1,-\frac{u_2}{2},\dots,-\frac{u_p}{p},0,0,0,\dots),\\
{\bf t}^{(2)}&\&={\bf t}_{\bf
v}:=(-v_1,-\frac{v_2}{2},\dots,-\frac{v_q}{q},0,0,0,\dots).
\eea
The partition function may therefore be expressed as
 \be
 \label{++2NMM}
Z_N^{2NMM}({\bf u}, n, m, {\bf v})  = N!\sum_{\lambda,\mu \atop
\ell(\lambda), \ell(\mu)\le N} g_{\lambda \mu }(N,n,m)
s_\lambda({\bf t}_{\bf u})s_\mu({\bf t}_{\bf v}),
 \ee
where
\be
 g_{\lambda \mu }(N,n,m)=\det\; (B_{h_i+n,h_j'+m})|_{i,j=1,\dots,N},\quad
 h_i=\lambda_i-i+N,\quad h'_i=\mu_i-i+N,
\ee
and $B_{jk}$ are given by (\ref{bimoments-BEH}).


\section{Direct derivation of Schur function expansions
 \label{direct}}

We now give an alternative derivation of formulae (\ref{ssss})-(\ref{-+'})  through a direct reduction of the multiple double integral to determinants
in the bimoments, using the well-known identity \cite{A}
 \be
 \int \prod_{a=1}^N d\mu(x_a,y_a) \det \phi_i(x_j) \det \psi_k(y_l),
= N! \det G, \qquad  1 \le i,j,k,l \le N.
\label{detidentity}
\ee
where $\{\phi_i, \psi_i\}_{i=1,\dots ,N}$
are an arbitrary set of pairs of functions whose products are
integrable with respect to the two variable measure $d\mu(x,y)$ on
some suitable domain and
 \be G_{ij} := \int d\mu(x,y) \phi_i(x)
\psi_j(y) \qquad 1\le i,j \le N.
 \ee
We also make use of the following form of the Cauchy-Littlewood identity
\cite{Mac}
 \be e^{\sum_{k=1}^\infty kt_k \tilde{t}_k} =
\sum_{\lambda}s_\lambda({\bf t})s_\lambda(\tilde{{\bf t}}),
\label{Cauchy}
 \ee
 where the sum is over all partitions $\lambda$.

For any finite set of variables $(x_1, \dots , x_N)$, we denote by
$[x]$ the infinite sequence of monomial sums
 \be [x]:=
(\sum_{a=1}^N x_a, {1\over 2}\sum_{a=1}^N x_a^2, \dots {1\over
j}\sum_{a=1}^N x_a^j, \dots),\quad
 [x^{-1}] : =
(\sum_{a=1}^N x_a^{-1}, {1\over 2}\sum_{a=1}^N x_a^{-2}, \dots
{1\over j}\sum_{a=1}^N x_a^{-j}, \dots).
 \ee
The identity (\ref{Cauchy}) may be used to express  the exponential factors
$e^{V(x, {\bf t}^{(1)})}$, $e^{V(y, {\bf t}^{(2)})}$,
$e^{V(x^{-1}, \bar{{\bf t})}^{(1)}}$, $e^{V(y^{-1}, \bar{{\bf
t}}^{(2)})}$ as series in Schur functions:
\bea
e^{V(x, {\bf t}^{(1)})} &\& =\sum_{\lambda}s_\lambda([x])s_\lambda({\bf
t}^{(1)}), \qquad
e^{V(y, {\bf t}^{(2)})} =\sum_{\lambda}s_\lambda([x])s_\lambda( {\bf t}^{(2)})
\label{expt}\\
e^{V(x^{-1},{ \bar{\bf t}}^{(1)})} &\&
=\sum_{\lambda}s_\lambda([x^{-1}])s_\lambda(\bar{{\bf t}}^{(1)}), \qquad
e^{V(y^{-1}, {\bar{\bf t}}^{(2)})}
=\sum_{\lambda}s_\lambda([x^{-1}])s_\lambda(\bar{{\bf t}}^{(2)}),
\label{expbart} \
\eea
where the sums are over partitions $\lambda$ with length $\ell(\lambda) \le N$.

Substituting the expansions (\ref{expt})  into the matrix integral
defining $Z_N({\bf t},n,m,\bar{\bf t}) $ gives
 \bea
Z_N({\bf t},n,m,\bar{\bf t}) &\&:= \prod_{a=1}^N\left(\int d\mu(x_a, y_a | {\bf t}, n, m,
\bar{\bf t})\right)  \Delta_N(x) \Delta_N(y)\cr
 &\& {\hskip -42 pt}=\prod_{a=1}^N
 \left(\int d\mu(x_a, y_a | {\bf 0}, n, m, \bar{\bf t})\right)
 \Delta_N(x) \Delta_N(y) \sum_\lambda s_\lambda({\bf t}^{(1)})
 s_\lambda([x]) \sum_\mu s_\mu({\bf t}^{(2)})s_\mu([y]). \cr
 &\&
 \eea
Making use of the Jacobi-Trudy formula
 \be
 \label{jacobi-trudy}
s_\lambda([x])
= {\det(x_i^{\lambda_j-j +N})\over \Delta_N(x)}, \quad s_\mu([x])
={ \det(y_k^{\lambda_l-l +N})\over \Delta_N(x)},
 \ee
 where
 \be
\lambda := \lambda_1 \ge \lambda_2 \ge \dots \ge
\lambda_{l(\lambda)}, \quad \mu := \mu_1 \ge \mu_2 \ge \dots \ge
\mu_{l(\mu)},
\ee
gives
 \be
 \label{doubleschursum}
 {\bf Z}_N = \sum_{\lambda}
\sum_{\mu}I^{nm}_{\lambda \mu} (\bar{\bf t})s_\lambda({\bf
t}^{(1)}) s_\mu({\bf t}^{(2)}),
 \ee
 where
 \be
I^{nm}_{\lambda\mu}(\bar{\bf t}):=\prod_{a=1}^N\left(\int
d\mu(x_a, y_a | {\bf 0}, n, m, \bar{\bf t})\right)
\det(x_i^{\lambda_j-j+N}) \det (y_k^{\mu_l-l+N}).
 \ee
Applying the identity (\ref{detidentity}) for
$\psi_i(x):=x^i$, $\phi_k(y):=y^k$ then gives
 \be
 \label{detGnmlamnbdamu}
I_{\lambda \mu}^{nm}
= N! \det G^{nm}_{\lambda \mu},
\ee
where
 \be
 \label{Gnmlamnbdamuij}
 (G^{nm}_{\lambda \mu})_{ij}:=
\int d\mu(x,y|{\bf 0}, n, m, \bar{\bf t}) x^{\lambda_i-i+N} y^{\mu_j -j +N} =
B_{n+h_i,m+h_j'}(0,0,\bar{\bf t}^{(1)},\bar{\bf t}^{(2)}),
 \ee
 proving the equality (\ref{++}).
 The other three cases of eqs. (\ref{--})-(\ref{-+}) are derived similarly.

We now derive the quadruple Schur function expansion (\ref{ssss}).
 To do this, we use  the product formula for characters (see \cite{Mac})
 \be s_\lambda([x])s_{\mu}([x])=\sum_{\alpha}
c_{\lambda\mu}^{\alpha}s_{\alpha}([x]),
 \ee
 where  the the Littlewood-Richardson coefficients $\{c_{\lambda\mu}^{\alpha}\}$ are
  the multiplicities with which the tensor representations
of type $\alpha$ appear in the tensor product of those of types $\lambda$ and
$\mu$. We have, as a consequence of the Jacobi-Trudy formula,
  \bea
s_\nu([x^{-1}])&\&=(x_1\cdots
x_N)^{-\nu_1}s_{\tilde\nu}([x]), \cr
s_\eta([y^{-1}])&\&=(y_1\cdots
y_N)^{-\mu_1}s_{\tilde\eta}([y])
 \eea
 where
$\tilde\nu=(\tilde\nu_1,\dots,\tilde\nu_N)$ and
$\tilde\eta=(\tilde\eta_1,\dots,\tilde\eta_N)$ and
 \be
\tilde\nu_i:=\nu_1- \nu_{N-i+1},\qquad
\tilde\eta_i:=\mu_1 - \eta_{N-i+1}.
 \ee
Combining this with
 \be
s_\lambda([x])s_{\tilde\nu}([x])=\sum_{\alpha}
c_{\lambda\tilde\nu}^{\alpha}s_{\alpha}([x]),\quad
s_\mu([y])s_{\tilde\eta}([y])=\sum_{\beta}
c_{\mu\tilde\eta}^{\beta}s_{\beta}([y]),
 \ee
gives
 \be
 \label{4schurexp}
Z_N({\bf t},n,m,\bar{\bf t})=\sum_{\lambda,\mu,\nu,\eta\atop
\ell(\lambda),\ell(\mu),\ell(\nu),\ell(\eta)\le
N}I_{\lambda \mu \nu \eta}(N,n,m)s_\lambda({\bf
t}^{(1)})s_\mu({\bf t}^{(2)})s_\nu(\bar{\bf
t}^{(1)})s_\eta(\bar{\bf t}^{(2)}) ,
 \ee
 where
 \be
 \label{Ilmne}
I_{\lambda \mu \nu \eta}(N,n,m):=\sum_{\alpha,\beta}
c_{\lambda\tilde\nu}^{\alpha}
c_{\mu\tilde\eta}^{\beta}J_{\alpha\beta}^{(n,m)}
 \ee
and
  \bea
J_{\alpha\beta}^{(n,m)}&\&:=N!\det(B_{l_i+n, l'_j+m})\vert_{1 \le i,j \le N}, \\
l_i:=\alpha_i-i+N-\nu_1,&\& \quad l_i':=\beta_i-i+N-\eta_1.
 \eea

\section{Summary and further related work}
We have given a new two component fermionic representation for $2N$-fold integrals of type (\ref{Z_N}) and used it both to show their relationship to the
two-component TL hierarchy, and to get double and quadruple Schur
function expansions as perturbation series in the deformation parameters,
generalizing analogous formulae earlier obtained in \cite{HO1, HO2} and  \cite{OS}.
These results were also derived through a ``direct'' method, based on standard determinantal identities and character formulae. In another work
 \cite{HO3} we show how to use this fermionic representation to reduce the
 evaluation of  multiple integrals of rational symmetric functions arising as determinantal correlation functions in matrix models, using  Wick's theorem, to determinantal expressions involving at most double integrals, and the associated biorthogonal polynomials. These results may also alternatively be derived through direct methods \cite{HO4}, based on standard determinantal identities and partial fraction expansions.  In \cite{HO5}  a fermionic representation for models of matrices coupled in a chain \cite{EM} is also derived.

\appendix
\section{Appendix: Relation to unitary, hermitian and  normal matrix models
  \label{AppendA}}
  \renewcommand{\theequation}{\Alph{section}-\arabic{equation}}
\setcounter{equation}{0}

In this appendix, we give a number of examples of matrix integrals that reduce to $2N$ fold integrals over their eigenvalues of the form ({\ref{Z_N}).
Two-matrix models involve integrals  of the type
\bea
< F >&\&= {1\over Z_N}\int F(M_1,M_2)d\Omega(M_1,M_2), \cr
Z_N  &\&= \int d\Omega(M_1, M_2)
\eea
where $F$ is some function of the entries of the matrices  $M_1$ and $M_2$
and $d\Omega(M_1, M_2)$ is some measure (possibly complex) on the set
of such pairs of matrices.
From among the various models that have been studied, we
review here three types: (A) normal matrix models, with spectrum supported in open regions of the complex plane; (B) models of pairs of
hermitian matrices or, more generally normal matrices with spectrum
supported on curve segments in the complex plane and (C) models of unitary matrices. The problem is to reduce the $\sim N^2$ integrations over the  independent matrix entries to just  $2N$ integrations over the eigenvalues of $M_1$ and $M_2$.  This is possible for certain choices of matrix measures.

\smallskip \noindent
(A) The simplest are models of normal $N \times N$ matrices $M$, where
\be
M_2^+=M_1=M,  \quad [M,M^+]=0,
\ee
 and the integrand is invariant under
$\Ub(N)$ conjugations. One can then diagonalize the matrix using such transformations,
\be
M \ra U M U^\dag,  \qquad U\in \Ub(N)
\ee
 and integrate over the group $\Ub(N)$ to reduce the integral
to a multiple integral over the eigenvalues with a
suitable domain of integration in the complex plane, and the induced
product measure multiplied by the factor $|\Delta_N(z)|^2 $ . The resulting reduced
measure (\ref{measure-defor-2cTL}) can be a rather general one
in the complex plane.

  For illustrative purposes, we consider the special case when the undeformed
  measure depends only on the combination $M  M^{\dag}$, and the
  deformed partition function  is of the form
  \be
  Z_N = \int dM\int dM^\dag (\det M)^n (\det M^\dag)^m
  e^{\Tr(\VV(M M^\dag) + \sum_{i=1}^\infty \left(t^{(1)}_i M^i + t^{(2)}_i (M^\dag)^i\right)}.
  \ee
   Diagonalizing by $\Ub(N)$ conjugation and integrating over the
   group $\Ub(N)$, this reduces, up to a proportionality constant $C_N$, to:
   \be
   Z_N = C_N \int dz_1 d\bar{z}_1 \cdots  \int dz_N d\bar{z}_N  |\Delta_N(z)|^2
   \prod_{a=1}^N z_a^n \bar{z}_a^me^{\VV(|z_a|^2) + \sum_{i=1}^\infty( t_i^{(1)} z_a^i + t_i^{(2)}\bar{z_a}^i)}.
   \ee
   Because of the polar rotational invariance in each complex $z_a$-plane, the angular parts of the bimoment integrals appearing in (\ref{Gnmlamnbdamuij})
   may be evaluated. Assuming, e.g. that $n\ge m$, the only nonvanishing
   terms  in the sum (\ref{doubleschursum}) are those
   for which the partitions $\lambda$ and $\mu$ are related by;
   \be
   \lambda_i +n =\mu_i +m
   \ee
   Evaluating the angular parts of the bimoment integrals, (\ref{doubleschursum})
  reduces to:
   \be
   Z_N = C_N \sum_{\lambda \atop \ell(\lambda)\le N} g_\lambda(n)
   s_{\lambda}({\bf t}^{(1)})   s_{\lambda +n-m}({\bf t}^{(2)})
   \ee
 where  $\lambda +n-m$ denotes the partition $(\lambda_1 +n-m, \cdots, \lambda_N+n-m)$ and
   \bea
   g_\lambda(n) &\&:=\pi^N \prod_{i=1}^N \left(\int_0^\infty e^{\VV(X)}
   X^{\lambda_i -i +n+N } dX\right).
   \eea

\noindent
(B) In the case where $(M_1,M_2)$ are a pair of independent  hermitian matrices,
or normal matrices with spectral supports along some specified curve segments,
the problem is more involved. The first example is the
 hermitian two-matrix model of Itzykson and Zuber \cite{IZ} (see also
\cite{Mehta}, where these are referred to as unitary ensembles). The partition
function is
\be
\label{I_N}
I_N=\int\int e^{Tr (V_1(M_1)+V_2(M_2))}e^{Tr
M_1M_2}d\Omega(M_1)d\Omega(M_2)
\ee
where
\be
d\Omega(M)=\prod_{i=1}^N dM_{ii} \prod_{i<j}^N  d\Re M_{ij}
\prod_{i<j}^N  d\Im M_{ij}.
\ee
The {\it potentials} $V_1, V_2$ can be fairly general, but most often
are taken as polynomials.
Diagonalizing the matrices $M_1$ and $M_2$ via two distinct
$\Ub(N)$ conjugations,
\be
M_i=U_iXU_i^{-1},\ i=1,2, \quad X:= \diag(x_1, \dots, x_N), \quad
Y:= \diag(y_1, \dots, y_N),
\ee
and integrating over $U_1$, one obtains
\be
\label{reduced_hermitian2M}
I_N =V_N\int\int dx_1 dy_1 \cdots
\int \int dx_N dy_N J_N(X, Y) \Delta_N(x)^2\Delta_N(y)^2\prod_{i=1}^N
 e^{Tr (V_1(x_i)+V_2(y_i))},
\ee
where
\be
J_N(X, Y) := \int_{\Ub(N)}e^{Tr UXU^{\dag}Y}d_*U
\ee
is the remaining integral over the unitary group, $d_*U$ is the normalized Haar measure and
\be
V_N := {(2\pi)^{N(N-1)} \over \left(\prod_{k=1}^N k!\right)^2}.
\ee
The observation of \cite{IZ} was that
\be
\int_{\Ub(N)} e^{Tr UXU^\dag Y} d_*U =\left(\prod_{k=1}^{N-1}k!\right)
{\det (e^{x_iy_j})\over\Delta_N(x)\Delta_N(y)}.
\ee
Using the anti-symmetry of determinants and changes of
variables, one then obtains the reduced integral
\be
I_N= {(2\pi)^{N(N-1)} \over \prod_{k=1}^N k!} \int\int dx_2 dy_1 \dots
\int\int dx_N dy_N
\Delta_N(x)\Delta_N(y)\prod_{i=1}^N
 e^{V_1(x_i)+V_2(y_i)}e^{x_iy_i} ,
\ee
which is proportional to  (\ref{Z_N}) for  measures  (\ref{measure-defor-2cTL})  of
the form
\be
\label{expbimeasure}
d\mu(x,y)=e^{V_1(x)+V_2(y)}e^{xy}dxdy.
\ee

This same computation is valid for any family of matrices that are
unitarily diagonalizable, even if the spectral support is not on
the real axis. Thus, if ($M_1, M_2$) are taken as normal matrices
with spectrum supported on some union of curves $\{\gamma_a\}$,  for $M_1$ and $\{\Gamma_b\}$ for $M_2$,
then the reduced form of the partition function is obtained by
replacing the double integrals  over the real axes in (\ref{reduced_hermitian2M})
by integrals of the form given, e.g., in eq.~(\ref{generalized2Mintegral}) for
the case of polynomial potentials (\cite{BEH1, BEH2}).

Generalizations of this construction were considered in \cite{ZJZ} and in
\cite{O1,OS}. We may replace the interaction term $e^{\Tr(M_1M_2)}$ in (\ref{I_N})
by a more general one of the form
 \be
 \label{tau_r}
\tau_r(N,M_1M_2):= \sum_{\lambda \atop  \ell(\lambda)\le N}
d_{\lambda,N} r_\lambda(N)s_\lambda(M_1M_2),
 \ee
where $\{r(j)\}_{j\in \Nb^+}$ is some sequence of complex numbers,
\be
\label{tau_r(1)}
r_\lambda(N):=\prod_{i,j\in \lambda}r(N+j-i),
\ee
with the product ranging over all nodes of the Young diagram $\lambda$ with
positions $(i,j)$,  and
\be
d_{\lambda, N} =s_\lambda(\Ib_N)
\ee
 is the dimension of the representation of ${\bf GL}(N)$ given by irreducible tensors of type $\lambda$.
 (We use here the abbreviated notation $\tau_r(N,M_1 M_2)$ to express what,
 in the more general setting of \cite{OS, O1, HO1}  was denoted
  $\tau_{r}(N, {\bf I}_N, M_1 M_2)$. Note that expressions such as (\ref{tau_r}),
viewed as functions of the trace invariants of the matrix $M_1M_2$,
are also KP $\tau$-functions.)
Here the Schur functions $s_\lambda(Z)$ are interpreted as conjugation invariant functions defined on the space of complex $N \times N$ matrices $Z$
(see Appendix \ref{AppendB}).
Then $J_N(X, Y)$ is replaced in eq.~(\ref{reduced_hermitian2M})  by the following more general integral
 (see  \cite{O1, OS} for details).
 \bea
J_{N,r}(X, Y)&\&:=\int_{\Ub(N)}
\tau_r(N,UXU^{-1}Y)d_*U
= \sum_{\lambda \atop \ell(\lambda)\le N} r_\lambda s_\lambda(X) s_\lambda(Y)\\
&\&=C_{N,r} \frac{\det (\tau_r(1,x_i y_j))}{\Delta_N(x)\Delta_N(y)},
\label{diagonalization}
 \eea
where
\be
\tau_r(1,x)= 1 + \sum_{k=1}^\infty r(1) \cdots r(k) x^k
\ee
and
\be
C_{N,r} := {1\over \prod_{k=1}^{N-1} \prod_{j=1}^k r(j)}.
\ee
 The reduced integral is again of the form (\ref{Z_N}) with the measure (\ref{expbimeasure}) replaced by
 \be
\label{tau_rbimeasure}
d\mu(x,y)=e^{V_1(x)+V_2(y)}\tau_r(1,xy)dxdy.
\ee
Note that, setting $Y=\Ib_N$ on the left hand side of eq.~(\ref{diagonalization}),
and taking the limits $\{y_j \ra 1\}_{j=1,\dots, N}$ on the right gives
\be
\label{tau_rderivdet}
\tau_r(N, X) = {C_{N,r} \over \prod_{k=1}^{N-1}k!}
{\det\left(x_i^{j-1} \tau_r^{(j-1)}(1, x_i)\right) \over \Delta_N(x)}.
\ee

Choosing
\be
\tau_r(1,x) = e^x, \quad r(j) = {1\over j}
\ee
we obtain from (\ref{tau_rderivdet})
\be
\tau_r(N, X) = e^{\tr(X)},
\ee
which is the Itzykson-Zuber case.

Another example is obtained by choosing
 \be
 \label{r_a_z}
\tau_r(1, x) = {1\over (1- z x)^{a-N+1}},  \qquad  r(j) = {z(a-N+j)\over j}
 \ee
 for some pair of constants $(a,z)$. In this case
 \be
 C_{N,r} = {\prod_{k=1}^{N-1}k! \over
 z^{{1\over 2}N(N-1)} \prod_{k=1}^{N-1} (a-N+1)_k}
 \ee
 where
 \be
  (a-N+1)_k:= \prod_{j=1}^{k}(a-N+j)
 \ee
 is the Pochhammer symbol, and
 \be
 \label{J_N_r_a_z}
 J_{N,r}(X,Y) = C_{N,r} {\det(1 - z x_i y_j)^{N-a-1}\over \Delta_N(x) \Delta_N(y)}.
 \ee
 Eq.~(\ref{tau_rderivdet}) therefore gives
  \be
\tau_r(N, X)=\det(\Ib_N-z X)^{-a}.
 \ee
 Thus the integral (\ref{reduced_hermitian2M}) is replaced by
 \bea
 \label{charpolycoupling}
I_{N,r}&\&=\int\int e^{Tr (V_1(M_1)+V_2(M_2))}\det(\Ib_N- z M_1
M_2)^{-a} d\Omega(M_1)d\Omega(M_2) \cr
&\&=\left ({2\pi^2 \over z}\right)^{{1\over 2}N(N-1)}{1\over \prod_{k=1}^{N-1} (a-N+1)_k}
 \int\int
\Delta_N(x)\Delta_N(y)\prod_{i=1}^N { e^{V_1(x_i)+V_2(y_i)}\over
(1 - z x_i y_i)^{a-N+1}}dx_idy_i,\cr
&\&
 \eea
 which is proportional to
(\ref{Z_N}) for  measures (\ref{measure-defor-2cTL})  of the
form
\be
 \label{charpoly measure} d\mu(x,y)=e^{V_1(x)+V_2(y)}
(1- z xy)^{N - a-1}dxdy.
\ee

\br For the case $a=0$ the right hand side of (\ref{J_N_r_a_z}),
up to the constant factor $C_{N,r}$, is
\be
 { \det((1-zx_iy_j)^{N-1})|_{i,j=1,\dots,N}\over
\Delta_N(x)\Delta_N(y)}, 
\ee 
which itself is a constant in the variables $\{x_i, y_i\}$, as of course 
it must be, since there is no coupling  in this case between the pairs 
of matrices $(M_1, M_2)$. This may be  seen by noting that  the numerator and
denominator are polynomials  of the same degree $N-1$  in all the
variables, and antisymmetric under the interchange of  any pair
$x_i \leftrightarrow x_j$ or $y_i \leftrightarrow y_j$. This is
therefore a rational function without poles, hence a polynomial,
which is bounded in all the variables, and therefore a constant.
\er

\noindent
(C) For unitary two-matrix models  \cite{OS} (referred to in \cite{Mehta}
as circular ensembles), $M_i\in \Ub(N),\ i=1,2$ ,
we have
 \bea
I_N^{gen}&\&=\int_{\Ub(N)}\int_{\Ub(N)}\ e^{Tr
(V_1(M_1)+V_2(M_2))}\tau_r(N,(M_1M_2)^{-1})d_*M_1d_*M_2 \cr
&\&=C_{N,r}V_N N! \oint\dots\oint
\Delta_N(x)\Delta_N(y)\prod_{i=1}^N
 e^{V_1(x_i)+V_2(y_i)}\tau_r(1,(x_i y_i)^{-1})\frac{dx_i}{x_i}\frac{dy_i}{y_i}
\eea
where $d_*M_1$, $d_*M_2$ are the Haar measure on  two copies of the
group $\Ub(N)$, and $x_i$ and $y_i$, $i=1,\dots,N$ are eigenvalues of the matrices
$M_1$ and $M_2$, with values on the unit circle in the complex $x_i$ and $y_i$ planes. This is related to  measures  in (\ref{measure-defor-2cTL})  of
the form
\be
d\mu =e^{V_1(x)+V_2(y)}\tau_r(1,(xy)^{-1}){dxdy \over xy}.
\ee
An example is the case
\bea
\tau_r(N,(M_1M_2)^{-1})&\&=e^{Tr (M_1M_2)^{-1} } \\
\tau_r(1,(x y)^{-1})&\&=e^{(xy)^{-1}},
\eea
which was considered in detail in \cite{ZJ, ZJZ}, where the large
$N$ limit of this model was also studied.

Another simple example, which is the unitary analog of the one above, is
 \bea
I_N^{gen}&\&=\int\int e^{Tr (V_1(M_1)+V_2(M_2))}\det(\Ib_N-z
(M_1M_2)^{-1})^{-a}d_*M_1d_*M_2 \cr &\&=C_{N,r}V_N N!
\oint\dots\oint \Delta_N(x)\Delta_N(y)\prod_{i=1}^N {
e^{V_1(x_i)+V_2(y_i)}\over (1- {z\over x_iy_i})^{(a-N+1)}}
\frac{dx_i}{x_i}\frac{dy_i}{y_i},
 \eea
which corresponds to  measures  in
(\ref{measure-defor-2cTL})  of the form \be d\mu
=e^{V_1(x)+V_2(y)}\left(1- {z\over xy}\right)^{N-a-1}{dxdy \over xy}. \ee

In this case,
$r$ is the same as (\ref{r_a_z})
 \be
 \label{rz_det}
 r(j) := {z(a-N+j )\over j},
 \ee
 and hence
 \bea
\tau_r(N, (M_1M_2)^{-1})&\&=\det(\Ib_N-z
(M_1M_2)^{-1})^{-a}  \\
\tau_r(1,{1\over xy}) &\&= {1\over (1- {z\over xy})^{a-N+1}}.
 \eea


\section{Appendix: derivation of formula (\ref{schurfunctionproduct})\label{AppendB}}

In this appendix, we recall some definitions about
Schur function (see \cite{Mac} for further details) and
 results from refs.\cite{JM, DJKM} that are needed in the
 derivation.

 A {\it partition} is a sequence of non-negative integers
in  weakly decreasing order:
\be
\label{partition}
\lambda = (\lambda_1, \lambda_2, \dots,\lambda_r,\dots )\ ,\quad
\lambda_1 \ge \lambda_2 \ge \dots \ge\lambda_r \ge \dots,
\ee
where we identify $(\lambda_1,\dots,\lambda_n)$ with
$(\lambda_1,\dots,\lambda_n,0)$ and, in the case of an infinite
sequence, assume $\lambda_r=0$ for $r\gg0$. The $\lambda_i$ in
(\ref{partition}) are called the {\it parts} of $\lambda$. The
number of nonzero parts is the  {\it length} of $\lambda$, denoted by
$\ell(\lambda)$.
We use the notation $\bt=(t_1,t_2,\dots)$ and
$\bar{\bt}=(\bar{t}_1,\bar{t}_2,\dots)$ in what follows.
The elementary Schur functions $\{s_i(\bt)\}_{i\in \Nb}$ are defined by
\be
\label{elemschur}
    \exp(V(x,\bt))=\sum_{k\ge0} s_i(\bt)x^k,
\ee
where
\be
V(x,\bt):=\sum_{k=1}^\infty t_k  x^k.
\ee
The Schur function $s_\lambda({\bf t})$ corresponding
to a partition $\lambda$ is given by
\be
\label{Schurt}
s_\lambda({\bf t})=\det\bigl(s_{\lambda_i-i+j}({\bf
t})\bigr)_{1\le i,j\le \ell(\lambda) } \ ,
\ee
where, for $k<0$, we put $s_k=0$ .
When the sequence $\{t_j\}_{j\in \Nb}$ is identified
with the elementary trace invariants $\{ {1\over j} \Tr(Z^j)\}_{j\in \Nb}$ of
elements $Z\in GL(n)$ or $Z\in U(n)$, the Schur function $s_\lambda({\bf t})$
is the trace (character) of the rank $\ell(\lambda)$ irreducible tensor representation
whose symmetries are given by the Young diagram associated
to the partition  $\lambda$. When interpreted in this way, as a class function
on the group $GL(n)$, its values are denoted $s_\lambda(Z)$, and this
may be extended, by continuity, to all complex $N \times N$ matrices $Z$.

Using the notations of subsections \ref{freefermi} and \ref{2compfermi} we define,
for $m\neq 0$,
\be
H_m:=\sum_{i\in\Zb}f_i\bar{f}_{i+m}.
\ee
It follows that
\be
\label{Hnfianticoms}
[H_m,f_i]=f_{i-m},\quad [H_m,\bar{f}_i]=-\bar{f}_{i+m},
\ee
and also that
\be
\label{Hmvac}
H_m|0\rangle=0,\quad m>0,
\ee
and hence
\be
\label{expHnvac}
e^{\sum_{m=1}^\infty H_mt_m}|0\rangle=|0\rangle.
\ee
For any $a\in \AA$, let
\bea
  a(\bt)&\&:=e^{\sum_{m=1}^\infty H_mt_m}ae^{-\sum_{m=1}^\infty H_mt_m
  }=\exp(\mathop{\mathrm{ad}}
  \sum_{m=1}^\infty H_mt_m)\;a, \\
   a(\bar{\bt})&\&:=e^{\sum_{m=1}^\infty H_{-m}\bar{t}_m}a
   e^{-\sum_{m=1}^\infty H_{-m}\bar{t}_m}=
   \exp(\mathop{\mathrm{ad}}
  \sum_{m=1}^\infty H_{-m}\bar{t}_m)\;a.
\eea
Using (\ref{Hnfianticoms}) it is easily verified  that
\bea
\label{eq:time evolution}
    {f}(x)(\bt)&\&=\exp(V(x,\bt)){f}(x),\qquad
    \bar{f}(y)(\bt)=\exp(-V(y,\bt))\bar{f}(y),
    \label{eq:time evolution bar} \\
      \label{eq:time evolution bar}
    {f}(x)(\bar{\bt})&\&=\exp(V(x^{-1},\bar{\bt})){f}(x),\quad
    \bar{f}(y)(\bar{\bt})=\exp(-V(y^{-1},\bar{\bt}))\bar{f}(y),
\eea
and hence
\bea
\label{eq:f(t)}
    {f}_i(\bt)&\&=\sum_{k\ge0} s_{k}(\bt){f}_{i-k},\qquad
    \bar{f}_i(\bt)=\sum_{k\ge0}s_{k}(-\bt)\bar{f}_{i+k}, \\
\label{eq:f(t)bar}
    {f}_i(\bar{\bt})&\&=\sum_{k\ge0} s_{k}(\bar{\bt}){f}_{i+k},\qquad
    \bar{f}_i(\bar{\bt})=\sum_{k\ge0}s_{k}(-\bar{\bt})\bar{f}_{i-k}.
\eea

Define the states
\be
\langle N|:=\langle 0|\bar{f}_0\cdots \bar{f}_{N-1}, \qquad \langle
-N|:=\langle 0|{f}_{-1}\cdots {f}_{-N}.
\ee
Now consider the integers $h_1>\cdots >h_N\ge 0$ related to a
partition $\lambda=(\lambda_1,\dots,\lambda_N)$ by
\be
h_i=\lambda_i-i+N.
\ee
It follows from (\ref{expHnvac}) that
\be
   \langle N|e^{\sum_{m=1}^\infty H_mt_m}{f}_{h_1}\cdots {f}_{h_N}|0\rangle
    =\langle N|{f}_{h_1}(\bt)\cdots {f}_{h_N}(\bt)|0\rangle.
\ee
By Wick's theorem and  eqs.~(\ref{eq:f(t)}), (\ref{eq:f(t)bar}) this is equal to
\be
\det\; (\langle 0|\bar{f}_{N-i}{f}_{h_j}(\bt)|0\rangle)
\;|_{i,j=1,\dots,N}=\det\; s_{\lambda_j-j+i}(\bt)
\;|_{i,j=1,\dots,N} = s_{\lambda}({\bf t}),
\ee
where the second equality follows from (\ref{Schurt}).
Thus we obtain
 \be\label{1}
\langle N|e^{\sum_{m=1}^\infty H_mt_m}{f}_{h_1}\cdots
{f}_{h_N}|0\rangle =s_\lambda(\bt).
 \ee
Similarly, we have
\bea
   \langle -N|e^{\sum_{m=1}^\infty H_mt_m}\bar{f}_{-h_1-1}\cdots
   \bar{f}_{-h_N-1}|0\rangle
    &\&=\det \; \langle 0|f_{i-N-1} \bar{f}_{-h_j-1}(\bt)|0\rangle \;|_{i,j=1,\dots,N} \cr
&\& = \det \; s_{\lambda_j-j+i}(-\bt)\;|_{i,j=1,\dots,N}\cr
&\&=s_\lambda(-\bt) . \label{2}
\eea

To prove eq.~(\ref{schurfunctionproduct}) we use  factorization,
(\ref{1}) and (\ref{2})
\bea
&\&\langle N,-N| e^{H^{(1)}({\bf t}^{(1)})- H^{(2)}({\bf t}^{(2)})}
f^{(1)}_{h_1}{\bar f}^{(2)}_{-h'_1-1}\cdots f^{(1)}_{h_N} {\bar
f}^{(2)}_{-h'_N-1}|0,0\rangle  \cr
&\&=
(-1)^{{1\over 2} N(N+1)} \langle N| e^{H^{(1)}({\bf t}^{(1)})}
f^{(1)}_{h_1}\cdots f^{(1)}_{h_N} |0\rangle\langle -N| e^{-
H^{(2)}({\bf t}^{(2)})} {\bar f}^{(2)}_{-h'_1-1}\cdots  {\bar
f}^{(2)}_{-h'_N-1}|0\rangle \cr
&\&\\
&\& =(-1)^{\frac 12 N(N+1)}s_\lambda(\bt^{(1)})s_\mu(\bt^{(2)}),
\eea
where
\be
h_i:=\lambda_i-i+N, \qquad h_i':=\mu_i-i+N.
\ee
Similarly, from
\be
\langle \pm N|H_{-m}=0
\ee
it follows that
\be
\langle \pm N|e^{\sum_{m=1}^\infty H_{-m}\bar{t}_m}=\langle \pm N|
\ee
We therefore have
\be
   \langle N|{f}_{N-h_N-1}\cdots {f}_{N-h_1-1}e^{-\sum_{m=1}^\infty H_{-m}\bar{t}_m}|0\rangle
    =\langle N|{f}_{N-h_N-1}(\bar{\bt})\cdots {f}_{N-h_1-1}(\bar{\bt})|0\rangle
\ee
By Wick's theorem this is equal to
\be
\det\; (\langle 0|\bar{f}_{i-1}{f}_{N-h_j-1}(\bar{\bt})|0\rangle)
\;|_{i,j=1,\dots,N}=\det\; s_{\lambda_j-j+i}(\bar{\bt})
\;|_{i,j=1,\dots,N} =s_\lambda(\bar{\bt}),
\ee
where the last equality follows from (\ref{eq:f(t)bar}). Thus
 \be\label{3}
 \langle N|{f}_{N-h_1-1}\cdots {f}_{N-h_N-1}
 e^{-\sum_{m=1}^\infty H_{-m}\bar{t}_m}|0\rangle
    =(-1)^{\frac12 N(N-1)}s_\lambda(\bar{\bt})
 \ee
Similarly,  we obtain
\bea
   \langle -N|\bar{f}_{h_N-N}\cdots
   \bar{f}_{h_1-N}e^{-\sum_{m=1}^\infty H_{-m}\bar{t}_m}|0\rangle
   &\& =\det \; \langle 0|{f}_{-i} \bar{f}_{h_j-N}(\bar{\bt})|0\rangle \;|_{i,j=1,\dots,N} \cr
&\& =
\det \; s_{\lambda_j-j+i}(-\bar{\bt})\;|_{i,j=1,\dots,N}=s_\lambda(-\bar{\bt}),\cr
&\&
\eea
and hence
 \be\label{4}
 \langle -N|\bar{f}_{h_1-N}\cdots
   \bar{f}_{h_N-N}e^{\sum_{m=1}^\infty H_{-m}\bar{t}_m}|0\rangle
      =(-1)^{\frac12 N(N-1)}s_\lambda(\bar{\bt})
 \ee

By factorization of the appropriate VEV of two-component
fermions into the product of VEV's of one component fermions,
eqns.~(\ref{1}), (\ref{2}), (\ref{3}) and (\ref{4}) then imply
 (\ref{schurfunctionproduct}), as well as  (\ref{schurfunctionproduct--}),
 (\ref{schurfunctionproduct+-}) and (\ref{schurfunctionproduct-+}).

 \br
 Note that all partitions $\lambda$ considered in this appendix
have length $\ell(\lambda)\le N$.
\er


\section{Appendix: Quadruple Schur function series for $\tau$-functions of
 two component Toda lattice and derivation of formula
(\ref{ssss})\label{AppendC}}

Using results from refs.\cite{JM, DJKM}  one proves (see e.g. \cite{HO1}) the formulae
 \be
 \label{C-1}
\langle N|e^{\sum_{k=1}^\infty H_{k}t_k}=\sum_{\lambda}
s_\lambda({\bf t})\langle N; \lambda|,\quad e^{\sum_{k=1}^\infty
H_{-k}t_k}|N\rangle =\sum_{\lambda} s_\lambda({\bf
t})|N;\lambda\rangle,
 \ee
where the sums range over all partitions and the Fock
vectors on the right hand sides are defined as follows
 \be\label{basis-Fvector-left}
\langle N; \lambda|:=a(N,\lambda)\langle N| \prod_{i=1}^k
\bar{f}_{N+\alpha_i} {f}_{N-\beta_i-1}
 \ee
 \be\label{basis-Fvector-right}
|N; \lambda\rangle :=a(N,\lambda)\prod_{i=1}^k\bar{f}_{N-\beta_i-1}
f_{N+\alpha_i} |N\rangle.
 \ee
 where
\be
a(N,\lambda)=(-1)^{\beta_1+\cdots+\beta_k+\frac 12N(N-1)}
\ee
Here $k$ is the number of diagonal nodes of the Young diagram corresponding
to $\lambda$, $\alpha_j$ is
the number of nodes in the $j$th row to the right of the $(j,j)$ diagonal
one and $\beta_j$ is the number of nodes in the column below it.
It follows that the sets of integers $\{ \alpha_i \}$, $\{ \beta_i \}$  are strictly decreasing  ($\alpha_1>\cdots >\alpha_k$) and ($\beta_1>\cdots >\beta_k$), and they uniquely determine $\lambda$. The partition $\lambda$ is expressed in Frobenius notation
 (see ref.\cite{Mac}) as:
\be
\label{frob}
\lambda=(\alpha_1,\dots,\alpha_k|\beta_1,\dots,\beta_k).
\ee
The transposed partition is denoted $\lambda^{\tr}$, which
in Frobenius notation is
 \be\label{Fr-tr}
  \lambda^{\tr}:=(\beta_1,\dots,\beta_k|\alpha_1,\dots,\alpha_k).
 \ee
From the Jacobi-Trudy formulae (\ref{jacobi-trudy}) it follows that
 \bea
 \label{f-Schur}
f(x_1)\cdots f(x_N)|0\rangle &\&=\sum_{\lambda \atop \ell(\lambda)\le
N }|N;\lambda\rangle s_\lambda([x])\Delta_N(x)
\label{bar-f-Schur} \\
\bar{f}(y_1)\cdots \bar{f}(y_N)|0\rangle &\&=\sum_{\lambda \atop
\ell(\lambda)\le N }|-N;\lambda\rangle
(-1)^{|\lambda|}s_{\lambda^{\tr}}([y])\Delta_N(y).
 \eea

In the two component setting, we use the following notations:
 \bea
&\&\langle N^{(1)}, N^{(2)};\lambda^{(1)},\lambda^{(2)}|  \cr
 &\&:=
a(N^{(1)},\lambda^{(1)})a(N^{(2)},\lambda^{(2)})\sum_{i=1}^{k^{(2)}}
 \langle
N^{(1)},N^{(2)}|\prod_{i=1}^{k^{(1)}}
\bar{f}_{N^{(1)}+\alpha_i}^{(1)} {f}_{N^{(1)}-\beta_i-1}^{(1)}
\prod_{i=1}^{k^{(2)}} \bar{f}_{N^{(1)}+\alpha_i}^{(2)}
{f}_{N^{(1)}-\beta_i-1}^{(2)}, \cr
&\&  \\
&\&| N^{(1)}, N^{(2)};\lambda^{(1)}, \lambda^{(2)}\rangle \cr &\&:=
a(N^{(1)},\lambda^{(1)})a(N^{(2)},\lambda^{(2)})
\prod_{i=1}^{k^{(1)}} \bar{f}_{N^{(1)}-\beta_i-1}^{(1)}
{f}_{N^{(1)}+\alpha_i}^{(1)} \prod_{i=1}^{k^{(2)}}
\bar{f}_{N^{(2)}-\beta_i-1}^{(2)}
{f}_{N^{(2)}+\alpha_i}^{(2)}|N^{(1)},N^{(2)}\rangle \cr &\&
\eea
where
\be
\lambda^{(j)}=(\alpha_1^{(j)},\dots,\alpha_{k^{(j)}}^{(j)}
|\beta_1^{(j)},\dots,\beta_{k^{(j)}}^{(j)}),\quad j=1,2.
 \ee
The vectors  $\{\langle N^{(1)}, N^{(2)};\lambda^{(1)}, \lambda^{(2)}|\}$ and
$\{|N^{(1)}, N^{(2)};\lambda^{(1)}, \lambda^{(2)}\rangle\}$ form  dual
orthonormal bases for the left and right Fock spaces:
\be
\langle N^{(1)}, N^{(2)};\lambda^{(1)},\lambda^{(2)}|
 M^{(1)}, M^{(2)}; \mu^{(1)},\mu^{(2)}\rangle=
 \delta_{N^{(1)},M^{(1)}}\delta_{N^{(2)},M^{(2)}}
 \delta_{\lambda^{(1)},\mu^{(1)}}\delta_{\lambda^{(2)},\mu^{(2)}}.
\ee

Using (\ref{C-1}) for each component, we then  have
 \bea
\label{N1N2lambda1lambda2left}
\langle N^{(1)},N^{(2)}|e^{\sum_{k=1}^\infty
H_{k}^{(1)}t_k^{(1)}-\sum_{k=1}^\infty
H_{k}^{(2)}t_k^{(2)}}&\&=\sum_{\lambda,\mu} s_\lambda({\bf
t}^{(1)})s_\mu(-{\bf t}^{(2)} )\langle
N^{(1)}+\lambda,N^{(2)}+\mu|, \cr
&\& \\
\label{N1N2lambda1lambda2right}
  e^{\sum_{k=1}^\infty
H_{-k}^{(2)}t_k^{(2)}-\sum_{k=1}^\infty
H_{-k}^{(1)}t_k^{(1)}}|N^{(1)},N^{(2)}\rangle
&\&=\sum_{\lambda,\mu} s_\lambda({\bf t}^{(1)})s_\mu(-{\bf
t}^{(2)} )|N^{(1)}+\lambda,N^{(2)}+\mu \rangle.
\cr
&\&
 \eea
This implies that each two-component TL $\tau$-function
 (\ref{2-comp-TL-tau}) may be expanded in  a quadruple Schur
 function series:
 \be
 \label{4Schur-2cTL}
 \langle N+n,-N-m|e^{H({\bf t})} ge^{\bar{H}(\bar{\bf t})}
|n,-m\rangle = \sum_{\lambda,\mu,\nu,\eta} g_{\lambda \mu \nu \eta}(N,n,m)s_\lambda({\bf
t}^{(1)})s_\mu(-{\bf t}^{(2)})s_\nu(\bar{\bf
t}^{(1)})s_\eta(-\bar{\bf t}^{(2)})
 \ee
where
 \be\label{matr-el-g}
g_{\lambda \mu \nu \eta}(N,n,m):=\langle
N+n+\lambda,-N-m+\mu|g|n+\nu,-m+\eta\rangle
 \ee
\br
The signs of the arguments of the Schur
functions on the right hand side of (\ref{4Schur-2cTL})
can be reversed using (see ref.\cite{Mac})
 \be
 \label{transposeschur}
s_{\lambda}({\bf t})=(-1)^{|\lambda|}s_{\lambda^{\tr}}(-{\bf t}).
 \ee
Here $|\lambda|$ denotes weight of
$\lambda$, which in the Frobenius notation equals
$k+\alpha_1+\cdots + \alpha_k+\beta_1+\cdots +\beta_k$.
\er
It follows from eq.~(\ref{result2}) that the partition function $Z_N({\bf t},n,m,{\bf t})$
is given by (\ref{4Schur-2cTL}).
\bea
Z_N({\bf t},n,m,{\bf t}) &\&= (-1)^{{1\over 2}N(N+1)+mN}
e^{\sum_{\alpha=1}^2 \sum_{k=1}^\infty k t_k^{(\alpha)}
\bar{t}_k^{(\alpha)}}  \cr
&\&  \qquad \times
\sum_{\lambda,\mu,\nu,\eta} g_{\lambda \mu \nu \eta}(N,n,m)s_\lambda({\bf
t}^{(1)})s_\mu(-{\bf t}^{(2)})s_\nu(\bar{\bf
t}^{(1)})s_\eta(-\bar{\bf t}^{(2)}),
\eea
where, by the Cauchy-Littlewood identity
\be
e^{\sum_{\alpha=1}^2 \sum_{k=1}^\infty k t_k^{(\alpha)}
\bar{t}_k^{(\alpha)}}
=\sum_{\lambda,\mu,\nu,\eta}s_\lambda({\bf t}^{(1)})s_\mu({\bf t}^{(2)})s_\nu(\bar{\bf
t}^{(1)})s_\eta(\bar{\bf t}^{(2)}).
\ee
Using (\ref{transposeschur}), and the product formula
 \be
 \label{multipl-Schur}
 s_\lambda({\bf t})s_{\mu}({\bf t})=\sum_{\alpha}
c_{\lambda\mu}^{\alpha}s_{\alpha}({\bf t}),
 \ee
where $c_{\lambda\mu}^{\alpha}$ are the
Littlewood-Richardson coefficients (see \cite{Mac}),  we obtain
 \be
 \label{4Schur-2cTL=Z_N}
Z_N({\bf t},n,m,\bar{\bf t})=\sum_{\lambda,\mu,\nu,\eta}I_{\lambda \mu \nu \eta}(N,n,m)s_\lambda({\bf
t}^{(1)})s_\mu({\bf t}^{(2)})s_\nu(\bar{\bf
t}^{(1)})s_\eta(\bar{\bf t}^{(2)}),
 \ee
 where
 \be
I_{\lambda \mu \nu \eta}(N,n,m)=(-1)^{{1\over 2}N(N+1) +mN}
N!{\hskip -10 pt}
\sum_{\lambda',\mu',\nu',\eta' \atop\lambda'',\mu'',\nu'',\eta''}
{\hskip -20 pt}(-1)^{|\mu'|+|\eta'|}
c_{\lambda'\lambda''}^{\lambda}c_{\mu'\mu''}^{\mu}
c_{\nu'\nu''}^{\nu}c_{\eta'\eta''}^{\eta}
g_{\lambda',\mu',\nu',\eta'}(N,n,m).
 \ee
Wick's theorem can then be used to evaluate the right hand side
 of (\ref{matr-el-g}) and express the result in terms of the bimoments.
The quadruple Schur expansion follows here from
 the fact that $Z_N({\bf t},n,m,\bar{\bf t})$ is essentially a $\tau$
 function for the two-component TL hierarchy.

Rather than detailing this calculation, we now give an alternative method of evaluating
 $I_{\lambda \mu \nu \eta}(N,n,m)$ that uses the particular form of $g$
in eq.~({\ref{el-g}),  and which directly yields the  vanishing of $I_{\lambda \mu \nu \eta}(N,n,m)$ if the length of
any of partitions $\lambda, \mu, \nu, \eta$ exceeds $N$.

Using (\ref{result2}) and (\ref{N1N2lambda1lambda2left}), we have
\bea
\label{Z_NdoubleschurApowerM}
Z_N({\bf t},n,m,\bar{\bf t})&\&= (-1)^{{1\over 2}N(N+1)}\langle N,-N|e^{H({\bf
t})} {A({\bf 0},n,m,\bar{\bf t})}^N|0,0\rangle \cr
&\& = (-1)^{{1\over 2}N(N+1)}\sum_{\lambda,\mu} s_\lambda({\bf t}^{(1)})s_\mu(-{\bf
t}^{(2)} )\langle N+\lambda,-N+\mu|{A({\bf 0},n,m,\bar{\bf t})}^N
 |0,0\rangle, \cr
 &\&
\eea
where
\be
\label{ApowerN}
{A({\bf 0},n,m,\bar{\bf t})}^N= \prod_{i=1}^N \int
f^{(1)}(x_i){\bar f}^{(2)}(y_i)x^n (-y)^m e^{V(x^{-1},\bar{\bf
t}^{(1)})+V(y^{-1},\bar{\bf t}^{(2)}) }d\mu(x_i,y_i).
\ee
It follows from (\ref{f-Schur}) and (\ref{bar-f-Schur}),  the sign count from
 interchanging the orders of fermionic operators, and the
orthogonality relations that
 \be
 \label{schurVdM2}
\langle N, -N; \lambda, \mu|\prod_{i=1}^N  f^{(1)}(x_i){\bar
f}^{(2)}(y_i)|0,0\rangle = (-1)^{{1\over 2}N(N+1)} s_\lambda([x])s_{\mu^{\tr}}([y])\Delta_N(x)\Delta_N(y),
 \ee
 which vanishes unless $\ell(\lambda), \ell(\mu^{\tr}) \le N$.
Substituting (\ref{ApowerN}), (\ref{schurVdM2}) into
(\ref{Z_NdoubleschurApowerM}),  using (\ref{transposeschur}) and changing
$\mu\to \mu^{tr}$ in the summation then gives
\bea
Z_N({\bf t},n,m,\bar{\bf t}) &\&= {\hskip -10 pt} \sum_{\lambda, \mu \atop \ell(\lambda), \ell(\mu)\le N}  {\hskip -10 pt}
s_\lambda({\bf t}^{(1)}) s_\mu({\bf t}^{(2)})
\left(\prod_{a=1}^N \int  d\mu(x_i, y_i) e^{V(x_i^{-1},\bar{\bf
t}^{(1)})+V(y_i^{-1},\bar{\bf t}^{(2)}) } \right)s_\lambda([x]) s_\mu([y]).\cr
&\&
\eea

At this point, the calculation becomes identical to the one
in section \ref{direct}. Using the Cauchy-Littlewood formula
\be
e^{V(x^{-1},\bar{\bf t}^{(1)}) }= \sum_{\nu\atop \ell(\nu)\le
N}s_\nu([x^{-1}])s_\nu(\bar{\bf t}^{(1)}),\qquad e^{V(y^{-1},\bar{\bf
t}^{(2)}) }=
 \sum_{\eta\atop
\ell(\eta)\le N}s_\eta([y^{-1}])s_\eta(\bar{\bf t}^{(2)}),
\ee
 the identities (see section \ref{direct})
  \be
s_\nu([x^{-1}])=(x_1\cdots x_N)^{-\nu_1}s_{\tilde\nu}([x]),\qquad
s_\eta([y^{-1}])=(y_1\cdots y_N)^{-\eta_1}s_{\tilde\eta}([y])
 \ee
 where
$\tilde\nu=(\tilde\nu_1,\dots,\tilde\nu_N)$,
$\tilde\eta=(\tilde\eta_1,\dots,\tilde\eta_N)$ and
 \be
\tilde\nu_i:=\nu_1 -\nu_{N-i+1},\quad
\tilde\eta_i:=\eta_1 -\eta_{N-i+1}
 \ee
and the product formulae
 \be
s_\lambda([x])s_{\tilde\nu}([x])=\sum_{\alpha}
c_{\lambda\tilde\nu}^{\alpha}s_{\alpha}(x),\quad
s_\mu([y])s_{\tilde\eta}([y])=\sum_{\beta}
c_{\mu\tilde\eta}^{\beta}s_{\beta}([y]),
 \ee
 we obtain
 \be
Z_N({\bf t},n,m,\bar{\bf t})=\sum_{\lambda,\mu,\nu,\eta\atop
\ell(\lambda),\ell(\mu),\ell(\nu),\ell(\eta)\le
N}I_{\lambda \mu \nu \eta}(N,n,m)s_\lambda({\bf
t}^{(1)})s_\mu({\bf t}^{(2)})s_\nu(\bar{\bf
t}^{(1)})s_\eta(\bar{\bf t}^{(2)}) ,
 \ee
 where
 \be
I_{\lambda \mu \nu \eta}(N,n,m)=\sum_{\alpha,\beta}
c_{\lambda\tilde\nu}^{\alpha}
c_{\mu\tilde\eta}^{\beta}J_{\alpha\beta}^{(n,m)}
 \ee
and, in terms of the bimoments $B_{ij}$,
 \bea
J_{\alpha\beta}^{(n,m)}&\&:=N!\det(B_{l_i+n, l'_j+m})\vert_{1 \le i,j \le N}, \cr
l_i:=\alpha_i-i+N-\nu_1,&\& \quad l_i':=\beta_i-i+N-\eta_1.
 \eea


\section*{Acknowledgements}
The authors  would like to thank T. Shiota and J. van de Leur for
helpful discussions, and  (A.O.)  thanks A. Odzievich for kind
hospitality during his stay in Bia{\l}ystok in June 2005, which
helped stimulate ideas leading to this work.

\end{document}